\documentclass{aa}
\usepackage[varg]{txfonts}
\usepackage{graphicx}
\usepackage{longtable}
\usepackage[version=3]{mhchem}
\bibpunct{(}{)}{;}{a}{}{,} 

\usepackage{color}

\begin{document}

\title{Do siblings always form and evolve simultaneously? Testing the coevality of multiple protostellar systems through SEDs}

\author{N. M. Murillo\inst{1,2}
\and E. F. van Dishoeck\inst{1,2}
\and J. J. Tobin\inst{1}
\and D. Fedele\inst{2,3}
}

\institute{Leiden Observatory, Leiden University, P.O. Box 9513, 2300 RA, Leiden, the Netherlands
        \and Max-Planck-Institut f\"{u}r extraterrestrische Physik, Giessenbachstra\ss e 1, 85748, Garching bei M\"{u}nchen, Germany
        \and INAF-Osservatorio Astrofisico di Arcetri, L.go E. Fermi 5, I-50125 Firenze, Italy}

\date{}

\abstract
{{Multiplicity is common in field stars and among protostellar systems. Models suggest two paths of formation: turbulent fragmentation and protostellar disk fragmentation.}}
{{We attempt to find whether or not the coevality frequency of multiple protostellar systems can help to better understand their formation mechanism. The coevality frequency is determined by constraining the relative evolutionary stages of the components in a multiple system.}}
{{Spectral energy distributions (SEDs) for known multiple protostars in Perseus were constructed from literature data. \textit{Herschel} PACS photometric maps were used to sample the peak of the SED for systems with separations $\geq$7$\arcsec$, a crucial aspect in determining the evolutionary stage of a protostellar system. Inclination effects and the surrounding envelope and outflows were considered to decouple source geometry from evolution. This together with the shape and derived properties from the SED was used to determine each system's coevality as accurately as possible. SED models were used to examine the frequency of non-coevality that is due to geometry.}}
{{We find a non-coevality frequency of 33 $\pm$ 10\% from the comparison of SED shapes of resolved multiple systems. Other source parameters suggest a somewhat lower frequency of non-coevality. The frequency of apparent non-coevality that is due to random inclination angle pairings of model SEDs is 17 $\pm$ 0.5\%. Observations of the outflow of resolved multiple systems do not suggest significant misalignments within multiple systems. Effects of unresolved multiples on the SED shape are also investigated.}}
{{We find that one-third of the multiple protostellar systems sampled here are non-coeval, which is more than expected from random geometric orientations. The other two-thirds are found to be coeval. Higher order multiples show a tendency to be non-coeval. The frequency of non-coevality found here is most likely due to formation and enhanced by dynamical evolution.}}


\titlerunning{Do siblings always form and evolve simultaneously?}
\authorrunning{Murillo et al.}

\maketitle

\section{Introduction}
\label{sec:intro}
Multiplicity is common in stars: 46\% of the solar-type field stars (\citealt{raghavan2010}) and more than 82\% of the O- and B-type stars (\citealt{chini2012}) are multiple stars.
Multiple stars are responsible for some of the more interesting phenomena in evolved stars, for example in the dust and gas shells of evolved stars \citep{maercker2012,decin2015}, phenomena such as type Ia supernovae (SNe), blue stragglers and cataclysmic variables that are generated through mass transfer in close binaries.
Multiples are also laboratories in which to test models of stellar physics and the products of star formation \citep{duchene2013}.

\citet{chen2013} and \citet{tobin2016} found that the frequency of multiplicity is highest for deeply embedded protostars and decreases to pre-main sequence and field stars in the separation range of 15 to 10000 AU. 
These authors used Submillimeter array (SMA) 1.3 mm and 850 $\mu$m archival data and Very Large Array (VLA) 8 mm and 1 cm observations, respectively.
However, these surveys are incomplete toward small separations ($<$15 AU for the VLA and $<$600 AU for the SMA), and the derived frequency should be considered a lower limit.
This clearly shows that stars are frequently born as multiple stellar systems.

While it is considered that fragmentation within the parent cloud is the mechanism through which multiples form, it is uncertain at which point in time and on what scale the fragmentation occurs.
Models suggest one of two paths: turbulent fragmentation of the core ($\geq$1600 AU scales, e.g., \citealt{offner2010}), or gravitational instability of the protostellar disk ($<$500 AU scales, e.g., \citealt{stamatellos2009a,kratter2010}). 
While some mechanisms are thought to produce coeval systems, turbulence can cause density enhancements that can lead to non-coevality in multiple protostellar systems.
Dynamical ejections of close binaires can, on the other hand, yield apparently non-coeval systems.

\begin{figure}
\includegraphics[width=\columnwidth]{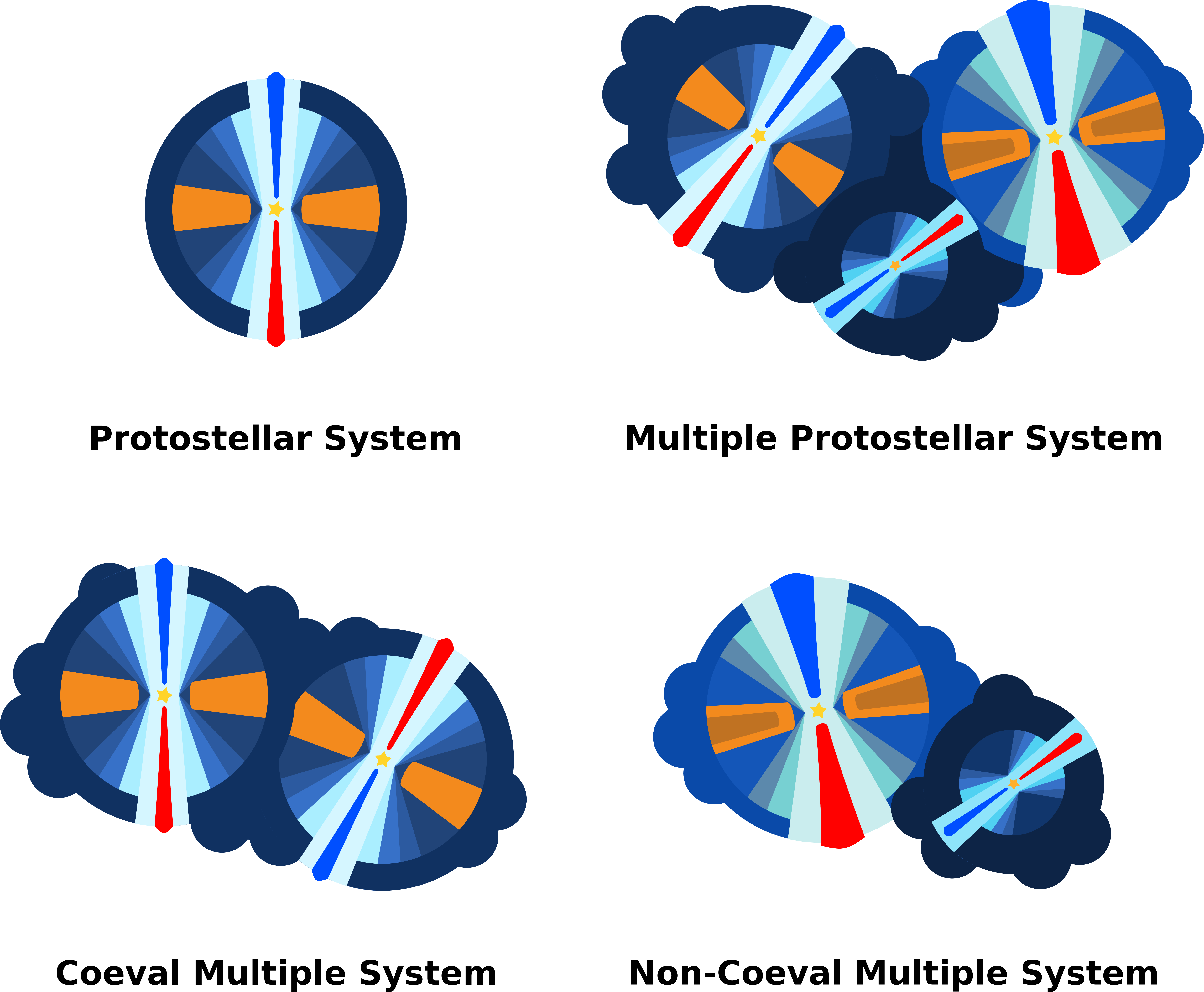}
\caption{Cartoon of the definitions used in this work. More evolved sources are represented by larger disks, wider outflow cavities and less envelope material}
\label{fig:cartoon}
\end{figure}

Early studies at disk scale ($\sim$100 AU) separations found
that 15\% out of 10 to 20 T Tauri and pre-main sequence binaries are formed of classical and weak-lined T Tauri stars, similar to mixed pairs in young binaries \citep{duchene1999,hartigan2003}. Classical T Tauri stars are generally considered to be younger and more actively accreting than weak-lined T Tauri stars \citep{duchene1999,kenyon1995}. Comparison of these binaries with isochrones showed that secondaries tend to be younger than primaries \citep{hartigan2003}, but it was suggested that this age difference would disappear with flatter isochrones.
A larger study of 65 T Tauri stars in Ophiuchus, Taurus and Corona Australis also found classical and weak-lined T Tauri binaries, in agreement with earlier studies, as well as Class I and II binaries \citep{mccabe2006} through comparison of color in K, L, [N] and 18 $\mu$m observations. This study noted that mixed pairs had a tendency of showing disks with low to no accretion, indicating different ages among the components, and supporting inside-out disk evolution.

\citet{kraus2009} studied the 36 known binaries in the Taurus-Auriga region ($d\sim$145 pc) with separations $>$200 AU, known spectral types and flux ratios with the aim to probe the coevality of pre-main sequence binaries.
Coevality of the sample of binaries was determined through comparison with a hybrid of two theoretical isochrones to estimate the ages of each component.
\citet{kraus2009} found that two-thirds of the pre-main sequence binaries are coeval with a dispersion lower than 1.4 Myr (0.16 dex), with no trend between age and mass or separation, suggesting that coevality is a product of formation.
It should be highlighted that only binaries were probed in \citet{kraus2009}, which raises the question of whether the coevality frequency is different when higher order multiples are considered. This is related to the dynamic evolution of multiple systems, since higher order multiples tend to disintegrate more readily \citep{reipurth2000}, and fewer of them survive to main sequence stages (11\% higher order multiples in solar-type stars, \citealt{raghavan2010}).

While isochrones are considered the best technique to determine ages, age determination is plagued by large uncertainties, bias and the assumptions made to estimate the age, namely the definition of $\tau$ = 0 \citep{soderblom2014}.
For embedded systems, determining the age is even more difficult due to the lack of information on the spectral type and stellar luminosity.
Using color, mass accretion rates and inner disk holes to determine evolutionary classfication and ages, while useful for T Tauri stars and even for a few Class I protostars \citep{duchene1999,hartigan2003,mccabe2006}, becomes difficult for deeply embedded sources, where near- infrared detections are often lacking and accretion can be more variable \citep{audard2014}. 
A more viable focus therefore is to probe the relative evolutionary stages of the components of multiple systems.
While the age coevality will not be probed, the evolutionary coevality, which sets the conditions for the system's life, will be probed and can provide insight into the question.

The evolutionary stage of protostars is usually defined by the spectral energy distribution (SED) shape, infrared spectral index $\alpha_{\rm IR}$, bolometric temperature $T_{\rm bol}$  and the ratio of submillimeter luminosity $L_{\rm submm}$ to bolometric luminosity $L_{\rm bol}$, which reflects the ratio of stellar mass $M_{*}$ to envelope mass $M_{\rm env}$ \citep{froebrich2005}.
The SED peak will tend to move toward the shorter wavelengths as the protostar evolves and disperses its envelope, changing the shape of the SED. 
It is expected that the parameters derived from the SED also reflect this, for example, $T_{\rm bol}$ will increase as the protostar sheds its natal cocoon and $\alpha_{\rm IR}$ will decrease as the protostar moves from the embedded phases to Class II.
As the envelope is dispersed, the submillimeter luminosity will decrease and therefore $L_{\rm submm}$ / $L_{\rm bol}$ will also decrease.
In-depth studies of some individual embedded multiple protostellar systems suggest non-coevality in embedded systems (such as L1448N A \& B: \citealt{ciardi2003}; NGC1333 SVS13: \citealt{chen2009}; L1448C: \citealt{hirano2010}; VLA1623: \citealt{murillo2013}) based on these criteria.
However, the geometry, or in other words, the inclination and outflow cavity of the observed protostellar system, affects the shorter wavelength ($\leq$ 70 $\mu$m) part of the SED. 
This in turn affects the derivation of parameters from the SED, some more strongly than others \citep{whitney2003,robitaille2006,crapsi2008}, and the evolutionary stage classification \citep{enoch2009,dunham2014}. 
Studies of modeled protostellar SEDs demonstrate that accurately constraining the inclination of the source provides more accurate estimates of the derived parameters and thus of the evolutionary stage classification \citep{offner2012b}.

The inclination of the protostellar system with respect to our line of sight can be estimated from outflow observations and is derived with more precision from rotationally supported disk structures, if present.
Protostellar systems alter their environment as they evolve, clearing out envelope material through widening of the outflow cavity \citep{arce2006}, accretion and concentration of material onto the protoplanetary disk.
Hence, the envelope and outflow can further constrain the evolutionary stage of the source through the chemical and physical structure of the envelope and core.
As a consequence, to establish the evolutionary stage of a source and eventually the coevality of a multiple protostellar system, the SED, derived properties, inclination and environment must be accounted for.

In this work we present the construction and analysis of the SEDs of all identified protostellar systems in the Perseus molecular cloud, the largest sample of Class 0, I and II. 
Perseus is the main target of this work because it is a well-studied region whose multiplicity and environment are relatively well known.
This provides data over a wide range of wavelengths and resolutions and both continuum and line emission towards most, if not all, of the region.

For this purpose, literature and archival data were used to construct the short ($<$70 $\mu$m) and long ($>$160 $\mu$m) wavelength regimes of the SEDs.
\textit{Herschel Space Observatory} Photodetector Array Camera and Spectrometer (PACS, \citealt{poglitsch2010}) photometric maps were used to cover the peak of the SEDs (70, 100 and 160 $\mu$m), without which large uncertainties arise in the parameters derived from the SED. The \textit{Herschel} PACS beamsize limits the range of component separations that can be probed to $\gtrsim$7$\arcsec$ , which at the distance of Perseus ($d \sim$ 235 $\pm$ 18 pc, \citealt{hirota2008,hirota2011}) becomes $\sim$1600 AU. 
Below this angular resolution, the fluxes of multiple systems are difficult to disentangle.
Therefore the coevality frequency, system alignment and properties derived in this work, as well as the environment, will provide constraints for fragmentation models only at core scales ($\geq$1600 AU).

In this paper we present the constructed SEDs of all identified and known systems in Perseus, with special focus on multiple protostellar systems. The SEDs are constructed from literature and \textit{Herschel} PACS data, compared with canonical SEDs for different stages from \citet{enoch2009}. With this work, we aim to determine the frequency of coevality in multiple protostellar systems to provide constraints for multiple protostar formation scenarios. Section~\ref{sec:def} defines the concepts used in this work. The data and sample studied in this paper, as well as the construction of SEDs and derivation of derived properties, are described in Sect.~\ref{sec:sampledata}. The results, including an analysis of unresolved SEDs, are given in Sect.~\ref{sec:analysis}. Finally, the discussion and conclusions are given in Sects.~\ref{sec:discuss} and ~\ref{sec:conc}, respectively.

\section{Definitions}
\label{sec:def}
For consistency and clarity of the terms used throughout this work, definitions of certain terms are provided in this subsection. An illustration of these definitions is shown in Fig.~\ref{fig:cartoon}.

\textit{Protostellar system} is defined as a source and its surrounding environment composed of a disk, an envelope and a bipolar outflow.

\textit{Multiplicity} or \textit{multiple} is used to refer to a system consisting of two or more components or sources, regardless of whether they are stars or protostars. The terms binary, triples and higher order multiples are thus implicitly merged into this term.

\textit{Multiple protostellar system} or simply \textit{multiple system} here refers to two or more protostellar sources composing one system. Multiple systems are generally observed to share a common envelope and, in some cases, a common disk. 
We assume that a group of protostars is gravitationally bound unless there evidence to the contrary, if they were observed to have a common envelope in single-dish observations.
An observed group of protostellar systems is considered a multiple when several observations and studies confirm its multiplicity through both continuum and molecular line emission.

\textit{Coevality} is taken to mean the relative evolutionary stages of the components that make up a multiple protostellar system, accounting for the SED, derived properties, inclination and environment. Environment is taken here to mean the outflows, surrounding envelope,and disk(s), if any. A multiple system whose components show similar evolutionary stages is considered coeval, while a multiple system with different evolutionary stages is referred to as non-coeval. 

\textit{Resolved multiple system} is a system with confirmed multiplicity with separations $\geq$ 7$\arcsec$ that can be resolved in the \textit{Herschel} PACS maps.

\textit{Unresolved multiple system} is a system with confirmed multiplicity down to 0.08$\arcsec$ and separations $<$ 7$\arcsec$ that cannot be resolved in the \textit{Herschel} PACS maps.

\begin{figure*}
\centering
\includegraphics[width=\textwidth]{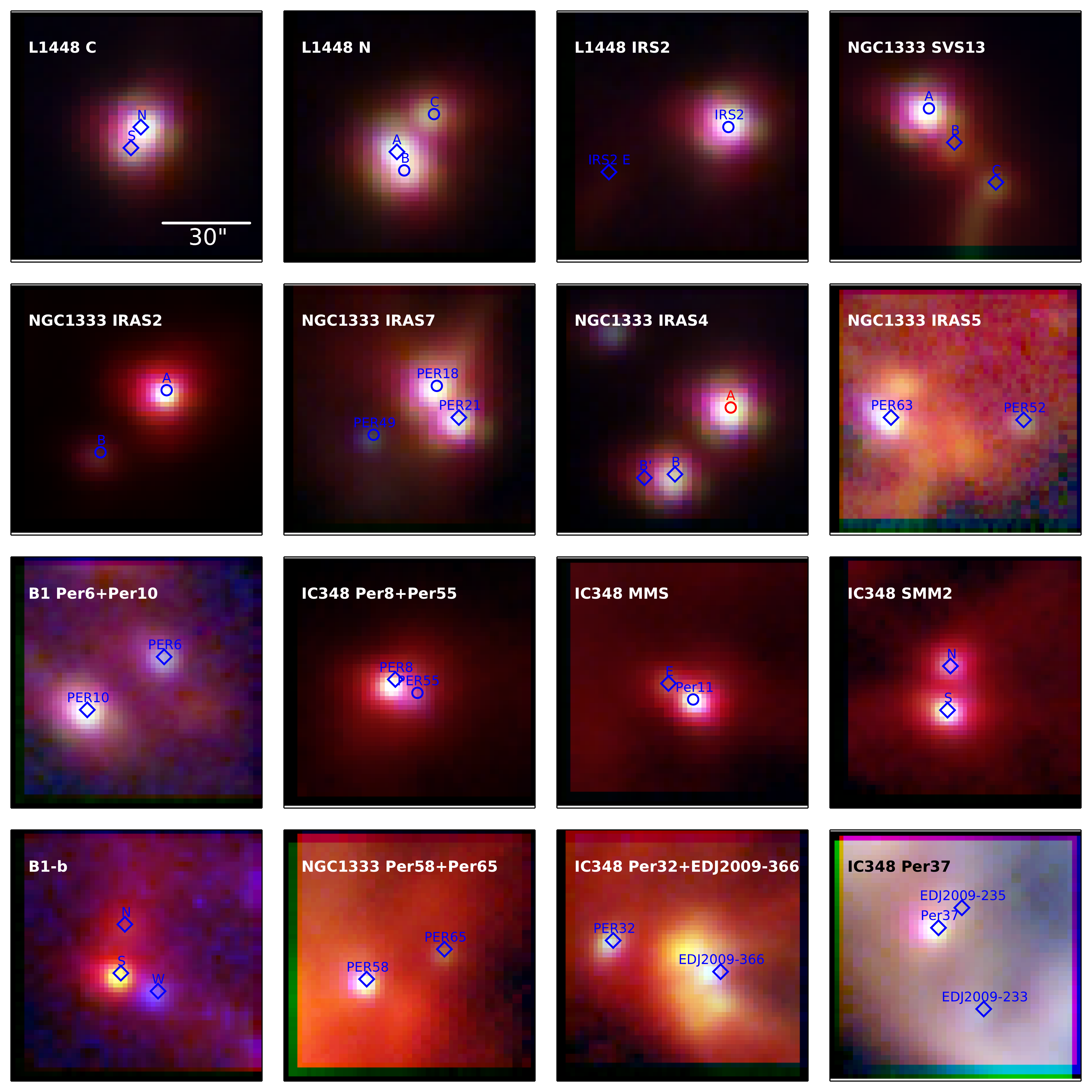}
\caption{\textit{Herschel} PACS stamps of resolved multiple protostellar systems in Perseus, except for NGC1333 IRAS4 which also shows IRAS4A, an unresolved protobinary. Each stamp is 80$\arcsec$ $\times$ 80$\arcsec$. 70, 100 and 160 $\mu$m are shown in blue, green and red, respectively. Blue symbols represent the components of a system, with circles denoting those with additional unresolved multiplicity and diamonds indicating those without (known) additional multiplicity. NGC1333 IRAS4A is marked in red. }
\label{fig:stamps}
\end{figure*}

\section{Sample and data}
\label{sec:sampledata}

\subsection{Source sample}
\label{sub:sample}
To study the coevality of multiple systems, the component protostellar systems must be identified.
Perseus was chosen because of the large number of embedded young stellar objects in a single cloud at d$<$300 pc.
Our source sample list and coordinates are obtained from \citet{tobin2016}, who identified multiple systems in Perseus down to 15 AU separations in the VLA Disk and Multiplicity survey of Perseus protostars (VANDAM) survey.
At the same time, the source sample was divided, based on the findings of \citet{tobin2016}, into three categories: resolved multiple systems and unresolved multiple systems with our 7$\arcsec$ separation, and single protostars. 
The source sample is listed in Tables~\ref{tab:resolved} and ~\ref{tab:unresolved}. 
Multiple systems and their components are referred to by their most common name.
Components with designations PerXX up to 66 are shorthand for Per-emb XX from \citet{enoch2009}.
Sources with designations EDJ2009-XXX refer to sources from \citet{evans2009}.
Duplicated systems in Tables~\ref{tab:resolved} and ~\ref{tab:unresolved} arise because some multiple systems have components that have been observed to be close binaries with separations $<$7$\arcsec$ (e.g., NGC1333 IRAS7 Per18, \citealt{tobin2016}). 
Half of the wide resolved multiple systems in our sample have a close unresolved companion, that is, 8 out of 16 systems. Of these 8 systems, 3 (L1448N, NGC1333 IRAS2 and NGC1333 IRAS7) have two resolved components with an unresolved companion each.
Confirmed single protostellar systems are listed and discussed in Appendix~\ref{app:singles}.

The sources in our sample have all been confirmed to be protostars through studies at multiple wavelengths, ruling out background sources, AGB stars, or galaxies.

\begin{table*}
\caption{Sample of resolved multiple protostellar systems (separation $\geq$ 7$\arcsec$)}
\label{tab:resolved}
\centering
\begin{tabular}{c c c c c}
\hline \hline
System & Component & RA (J2000)\tablefootmark{a} & Dec. (J2000)\tablefootmark{a} & Separation\tablefootmark{b} ($\arcsec$) \\
\hline
L1448 C & N & 03:25:38.87 & +30:44:05.40 & ... \\
... & S & 03:25:39.14 & +30:43:58.30 & 8.1 \\
L1448 N & A & 03:25:36.53 & +30:45:21.35 & ... \\
... & B & 03:25:36.34 & +30:45:14.94 & 7.3 \\
... & C & 03:25:35.53 & +30:45:34.20 & 16.3 \\
L1448 IRS2 & IRS2 & 03:25:22.40 & +30:45:12.00 & ... \\
... & IRS2E & 03:25:25.66 & +30:44:56.70 & 46.9 \\
NGC1333 SVS13 & A & 03:29:03.75 & +31:16:03.76 & ... \\
... & B & 03:29:03.07 & +31:15:52.02 & 14.9\\
... & C & 03:29:01.96 & +31:15:38.26 & 34.7\\
NGC1333 IRAS2 & A & 03:28:55.57 & +31:14:37.22 & ... \\
... & B & 03:28:57.35 & +31:14:15.93 & 31.4 \\
NGC1333 IRAS7 & Per18 & 03:29:11.26 & +31:18:31.08 & ... \\
... & Per21 & 03:29:10.67 & +31:18:20.18 & 13.3\\
... & Per49 & 03:29:12.96 & +31:18:14.31 & 27.5 \\
NGC1333 IRAS4 & B & 03:29:12.01 & +31:13:08.10 & ... \\
... & B' & 03:29:12.83 & +31:13:06.90 & 10.6\\
NGC1333 IRAS5 & Per52 & 03:28:39.72 & +31:17:31.89 & ... \\
... & Per63 & 03:28:43.28 & +31:17:32.90 & 45.7 \\
B1 Per6+Per10 & Per6 & 03:33:14.40 & +31:07:10.88 & ... \\
... & Per10 & 03:33:16.45 & +31:06:52.49 & 31.9\\
IC348 Per8+Per55 & Per8 & 03:44:43.94 & +32:01:36.09 & ... \\
... & Per55 & 03:44:43.33 & +32:01:31.41 & 9.6\\
IC348 MMS & Per11 & 03:43:57.06 & +32:03:04.60 & ... \\
... & E & 03:43:57.73 & +32:03:10.10 & 10.2 \\
IC348 SMM2 & S & 03:43:51.08 & +32:03:08.32 & ... \\
... & N & 03:43:51.00 & +32:03:23.76 & 16.1 \\
B1-b & S & 03:33:21.30 & +31:07:27.40 & ... \\
... & N & 03:33:21.20 & +31:07:44.20 & 17.4\\
... & W & 03:33:20.30 & +31:07:21.29 & 13.9  \\
NGC1333 Per58+Per65 & Per58 & 03:28:58.44 & +31:22:17.40 & ... \\
... & Per65 & 03:28:56.31 & +31:22:27.80 & 28.9 \\
IC348 Per32+EDJ2009-366 & Per32 & 03:44:02.40 & +32:02:04.89 & ... \\
... & EDJ2009-366 & 03:43:59.44 & +32:01:53.99 & 36.6 \\
NGC1333 PER37 & Per37 & 03:29:18.89 & +31:23:12.89 & ... \\
... & EDJ2009-235 & 03:29:18.259 & +31:23:19.758 & 10.6 \\
... & EDJ2009-233 & 03:29:17.675 & +31:22:44.922 & 33.7 \\
\hline
\end{tabular}
\tablefoot{
\tablefoottext{a}{Coordinates from \citet{tobin2016}.}
\tablefoottext{b}{Separations are obtained from \citet{tobin2016} and are listed relative to the first component tabulated. Typical uncertainties in position are $<$0.1$\arcsec$.}
}
\end{table*}

\begin{table*}
\caption{Sample of unresolved multiple protostellar systems (separation $<$ 7$\arcsec$)}
\label{tab:unresolved}
\centering
\begin{tabular}{c c c c}
\hline \hline
System & RA (J2000) & Dec. (J2000) & Sepearation\tablefootmark{a} ($\arcsec$)\\
\hline
NGC1333 IRAS4A & 03:29:10.51 & +31:13:31.01 & 1.828\\
IRAS 03292+3039 & 03:32:17.95 & +30:49:47.60 & 0.085\\
IRAS 03282+3035 & 03:31:21.00 & +30:45:30.00 & 0.098\\
NGC1333 IRAS2A & 03:28:55.57 & +31:14:37.22 & 0.619 \\
NGC1333 IRAS2B & 03:28:57.35 & +31:14:15.93 & 0.311 \\
NGC133 IRAS7 Per18 & 03:29:11.26 & +31:18:31.08 & 0.081\\
NGC1333 IRAS7 Per49 & 03:29:12.96 & +31:18:14.31 & 0.313\\
L1448N C & 03:25:35.53 & +30:45:34.20 & 0.251\\
L1448N B & 03:25:36.34 & +30:45:14.94 & 0.226\\
Per17 & 03:27:39.09 & +30:13:03.00 & 0.273 \\
IC348 MMS Per11 & 03:43:57.06 & +32:03:04.60 & 2.950\\
NGC1333 SVS13A & 03:29:03.75 & +31:16:03.76 & 0.3\\
L1448 IRS2 & 03:25:22.40 & +30:45:12.00 & 0.751\\
L1455 FIR2 & 03:27:38.23 & +30:13:58.80 & 0.346\\
B1-a & 03:33:16.66 & +31:07:55.20 & 0.391\\
EDJ2009-269 & 03:30:43.91 & +30:32:46.28 & 0.539\\
IC348 Per55 & 03:44:43.33 & +32:01:31.41 & 0.613\\
EDJ2009-183 & 03:28:59.32 & +31:15:48.14 & 1.022 \\
L1448 IRS1 & 03:25:09.54 & +30:46:21.96 & 1.424\\ 
NGC1333 IRAS1 & 03:28:37.00 & +31:13:27.00 & 1.908 \\ 
EDJ2009-156 & 03:28:51.11 & +31:18:15.41 & 3.192 \\
Per32 & 03:44:02.40 & +32:02:04.89 & 5.910 \\
HH211 & 03:43:56.80 & +32:00:50.21 & 0.3\tablefootmark{b} \\
Per62 & 03:44:12.98 & +32:01:35.40 & 0.121\tablefootmark{c}\\
\hline
\end{tabular}
\tablefoot{
\tablefoottext{a}{Separations obtained from \citet{tobin2016}. Typical uncertainties in position are $<$0.1$\arcsec$.}
\tablefoottext{b}{The companion reported in \citet{lee2009} appears to be substellar with an orbital period of 3000 yr, which could explain the jet precession. \citet{lee2010} also argued for a third component at $<$30 AU, proposing that it is a very low-mass system.}
\tablefoottext{c}{Possibly unresolved binary.}
}
\end{table*}

\subsection{Literature data}
\label{sub:lit}
Most star forming regions have been observed at infrared and (sub)millimeter wavelengths at different epochs and with varying resolutions.
The first step to constructing SEDs is therefore a search of the available data in the literature.
It needs to be noted, however, that even though there is much information in the literature, not all protostellar systems have been homogeneously observed or photometry reported, making it impossible to have all SEDs sampled at the same wavelengths.

The near- to mid-infrared regime of protostellar SEDs is well characterized from 2MASS and \textit{Spitzer Space Telescope} observations with fluxes shortwards of 70 $\mu$m. 
The c2d catalog \citep{dunham2015} provides fluxes from 1.25 $\mu$m to 24 $\mu$m. The \textit{Spitzer} 70 $\mu$m fluxes are not considered here given the large beam and saturation of the MIPS instrument for the 70 $\mu$m detector and the superior quality of the \textit{Herschel} data.
Sensitivity limits at each wavelength from the respective instruments are taken as upper limits for sources that lacked an entry in the c2d catalog.
For NGC1333, integrated fluxes at wavelengths $<$70 $\mu$m were obtained from the compiled catalog of \cite{rebull2015} after conversion from magnitude to mJy units.

Submillimeter and millimeter integrated fluxes were collected from diverse interferometric continuum surveys (e.g.,  \citealt{looney2000,jorgensen2007,chen2013,yen2015}) as well as works reporting fluxes for individual protostellar systems (e.g., \citealt{chen2009,hirano2010,palau2014}).
Careful selection of the fluxes from literature was made to ensure that as much emission could be recovered from the observations
as possible, while at the same time the individual sources could be clearly and easily separated.
The VANDAM survey \citep{tobin2016} provides fluxes from 8 mm to 1 cm for all sources in the Perseus star forming region.
Interferometric observations are preferred over single-dish observations because of the resolution needed to separate the flux contribution from each component in a multiple system.
The typical fraction of recovered flux varies by telescope configuration, sensitivity and structure being probed.  \citet{tobin2015} provided a comparison that gives an idea of the recovered flux in interferometric observations.

Although data from the literature can cover the near- to mid-infrared and (sub)millimeter regimes of the SED, the peak of the SED is not well sampled typically at 70 to 160 $\mu$m. 
The lack of a well sampled SED peak can seriously underestimate the derived parameters and evolutionary classification of a protostar, and in turn the coevality determination of a system.
\textit{Herschel} PACS data are therefore crucial to this work.

\begin{table*}
\caption{\textit{StarFinder} photometry parameters}
\label{tab:SFp}
\centering
\begin{tabular}{c c c c c c c}
\hline \hline
Regions & Wavelength & Beam size\tablefootmark{a} & PSF aperture\tablefootmark{b} & FWHM\tablefootmark{c} & Deblend\tablefootmark{d} & Flux uncertainty \\
 & $\mu$m & $\arcsec$ & $\arcsec$ & & & mJy \\
\hline
NGC1333, B1, IC348 & 70 & 9.6 & 13.0 & 1.0 & N & 20 - 30 \\
... & 100 & 7.2 & 8.0 & 1.0 & N & 7 - 10 \\
... & 160 & 12.8 & 13.0 & 1.0 & N & 10 - 15 \\
L1448, L1455 & 70 & 9.6 & 13.0 & 0.7 & Y & 24 - 34 \\
... & 100 & 7.2 & 8.0 & 0.7 & Y & 8 - 16 \\
... & 160 & 12.8 & 13.0 & 0.7 & Y & 11 - 20 \\
\hline
\end{tabular}
\tablefoot{
\tablefoottext{a}{Measured from the FWHM of the extracted PSF.}
\tablefoottext{b}{This refers to the mask applied to the PSF
to include the PSF sidelobes.}
\tablefoottext{c}{Parameter to determine the smallest separation between close sources in terms of the FWHM.}
\tablefoottext{d}{Switch parameter to set whether detected sources are deblended.}
}
\end{table*}

\subsection{\textit{Herschel} PACS photometric maps}
\label{sub:herschel}
Archival photometric maps from \textit{Herschel} PACS from the Gould Belt Survey \citep{andre2010,pezzuto2012} were obtained from the Herschel Science Archive for the entire Perseus region.
The maps made with JScanmap were selected for performing the photometry (see Appendix~\ref{app:maps}).
From these data we can extract 70~$\mu$m $\leq$ $S_{\rm \nu}$ $\leq$ 160~$\mu$m integrated fluxes. Due to the resolution of \textit{Herschel} PACS observations, fluxes from each component in a multiple protostellar system can be extracted only for systems whose projected separations are $\geq$7$\arcsec$.

Star-forming regions tend to be clustered, hence, crowded-field photometry techniques are employed to best exploit the \textit{Herschel} PACS maps.
While aperture photometry is a simple and straightforward method, it is not a viable solution for crowded protostellar fields.
Point spread function (PSF) photometry presents a better solution to the problem at hand.
The IDL-based program \textit{StarFinder} \citep{diolaiti2000} was employed to perform photometry on the \textit{Herschel} maps.
The PSF was extracted from the maps themselves to account for the specific observation mode, which cannot be achieved as easily with modeled ideal PSFs.
Single isolated sources were used to extract the PSF, with moderate brightness, thus avoiding spikes and negative spots, and little to no surrounding nebulosity.
The extracted PSFs provide beam sizes of 9.6$\arcsec$, 7.2$\arcsec$  and 12.8$\arcsec$ for 70, 100 and 160 $\mu$m, respectively.
\textit{StarFinder} allows deblending of sources and setting a lower limit for the FWHM for source separation, which proves to be very useful in separating multiple systems from PACS maps.

Postcard maps of each sub-region of Perseus, measuring 44' $\times$ 44', were extracted for ease of photometry.
For postcard maps from the same larger map, the same PSF was used, which means that we required the PSF to be extracted only once per map using the best single-source targets.
To avoid an overestimation of the measured fluxes and facilitate source deblending, the extracted PSF was then masked by introducing an aperture factor.
The \textit{StarFinder} parameters for the photometry used in this work for each subregion and the typical flux uncertainty per wavelength are listed in Table~\ref{tab:SFp}. 
After PSF photometry was performed, PSF aperture and background corrections were applied to the raw fluxes.
The values used for aperture correction are tabulated in \cite{balog2014}.
A detailed explanation of the photometry with \textit{StarFinder} is given in Appendix~\ref{app:maps}.

\begin{figure*}
\centering
\includegraphics[width=\textwidth]{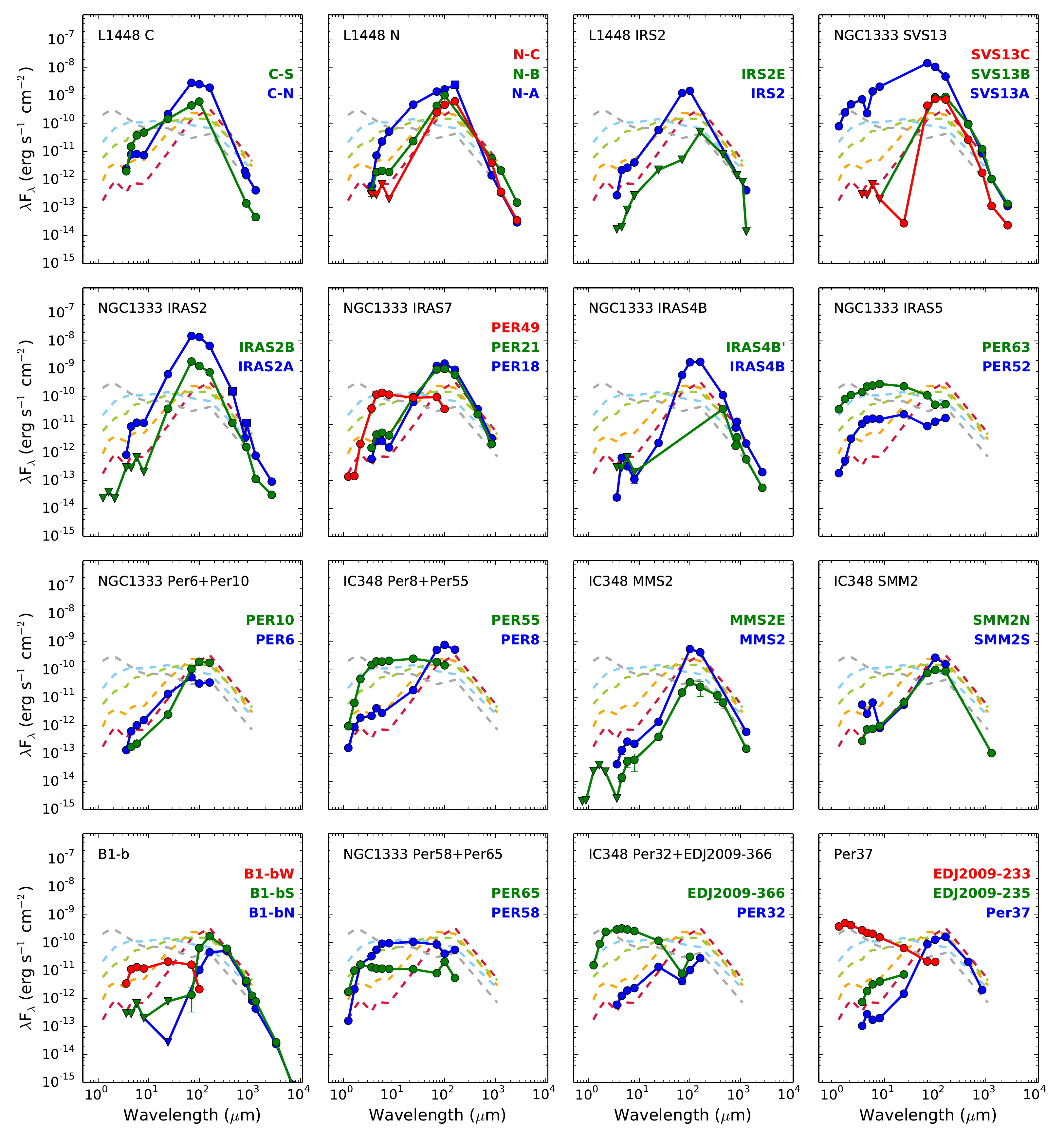}
\caption{Constructed SEDs for resolved multiple protostellar systems. Filled circles denote the fluxes without contamination from nearby sources. Triangles indicate upper limits. Each SED is overlaid with the template SEDs from \citet{enoch2009} for comparison (dashed lines).}
\label{fig:SEDresolved}
\end{figure*}

\begin{figure*}
\includegraphics[width=\textwidth]{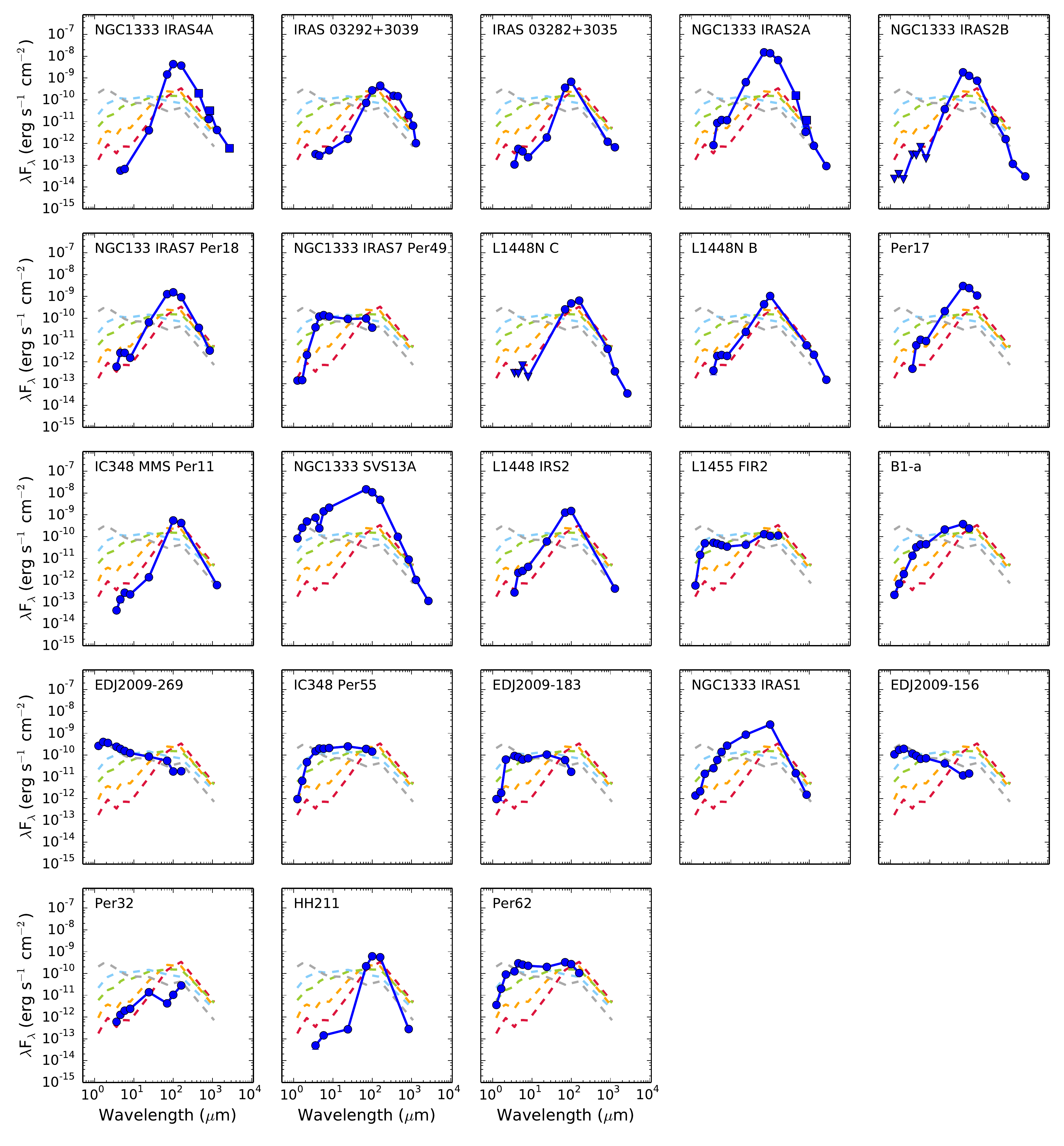}
\caption{Constructed SEDs for unresolved multiple protostellar systems. Other details are the same as in Fig.~\ref{fig:SEDresolved}.}
\label{fig:SEDunresolved}
\end{figure*}

\begin{table*}
\caption{Statistics from the constructed SEDs.}
\label{tab:SEDstats}
\centering
\begin{tabular}{c c c c c c}
\hline \hline
 & L1448 \& L1455 & NGC1333 & IC348 & B1 & Total \\
\hline
Multiple total & 9 & 17 & 9 & 5 & 40 \\
Resolved system & 3 & 8 & 4 & 1 & 16 \\
Unresolved system & 6 & 9 & 5 & 4 & 24 \\
Single & 7 & 20 & 9 & 12 & 48 \\
Total systems & 16 & 36 & 18 & 17 & 88 \\
Multiplicity frequency\tablefootmark{a} & 56.3\% & 47.2\% & 50.0\% & 29.4\% & 45.5\% \\
\hline
Resolved multiples: Coeval & 3 & 6 & 2 & 2 & 14 \\
Resolved multiples: Non-coeval & 1 & 3 & 2 & 1 & 7 \\
Total determined systems & 4 & 7 & 4 & 3 & 21 \\
Non-coevality frequency\tablefootmark{a} & 25\% & 29\% & 50\% & 66\% & 33 $\pm$ 10\% \\
\hline
\end{tabular}
\tablefoot{
\tablefoottext{a}{Calculated as the ratio of multiples or non-coeval systems over the total number of systems.}} 
\end{table*}

\subsection{SED construction}
\label{sub:sed}
Given the multiple names and identifiers each source has accumulated through surveys and literature, fluxes at different wavelengths were matched by means of the coordinates with a search radius of $\lesssim$ 4.5$\arcsec$.
The search radius was selected to be below the resolution limit of \textit{Herschel} and similar to the FWHM lower limit for \textit{StarFinder}, avoiding any confusion in source identification. 
Coordinates were obtained from (sub)millimeter interferometric observations given the higher angular resolution and because the source positions at these wavelengths are less likely to be contaminated by foreground stars (e.g., NGC1333 IRAS2B at $\lambda \lesssim$ 8 $\mu$m, \citealt{rodriguez1999}) or scattered light.

Care was taken that fluxes at all available wavelengths for each SED were separated from the other protostars in their system.
At 160 $\mu$m this criteria breaks down for systems with separations smaller than 9 to 10$\arcsec$. In these cases, the flux is flagged as combined and is noted in the plotted SEDs.
Upper limits are also flagged and noted with a different symbol in the plots.

L1448 IRS1, an unresolved multiple systems, has fewer than three points in the SED.
The same situation occurs for 7 single protostellar sources, listed in Table~\ref{tab:single}. 
Hence, these systems are not shown in Figs.~\ref{fig:SEDunresolved} and~\ref{fig:SEDsingle}.

\begin{table*}
\caption{SED derived properties}
\label{tab:param}
\centering
\begin{tabular}{c c c c c c c}
\hline \hline
System & Source & $T_{\rm bol}$ & $L_{\rm bol}$ & $L_{\rm fir}$ / $L_{\rm bol}$ & $L_{\rm submm}$ / $L_{\rm bol}$ & $\alpha_{\rm 3-24 \mu m}$ \\
& & K & $L_{\odot}$ & & & \\
\hline
L1448 C & N & 49.5 $\pm$ 15.8 & 10.27 $\pm$ 1.59 & 0.479 $\pm$ 0.104 & (7.7 $\pm$ 1.9)E-5 & 2.19 $\pm$ 0.38\\
… & S & 82.6 $\pm$ 12.1 & 1.98 $\pm$ 0.31 & 0.416 $\pm$ 0.092 & (3.2 $\pm$ 0.9)E-5 & 1.86 $\pm$ 0.65\\
L1448 N & A & 57.3 $\pm$ 9.8 & 8.08 $\pm$ 1.27 & 0.566 $\pm$ 0.124 & (9.4 $\pm$ 2.4)E-5 & 3.09 $\pm$ 0.72\\
… & B & 47.0 $\pm$ 25.7 & 2.11 $\pm$ 0.33 & 0.632 $\pm$ 0.137 & (1.8 $\pm$ 0.4)E-3 & 1.87 $\pm$ 0.37\\
… & C & 43.3 $\pm$ 0.2 & 2.89 $\pm$ 0.44 & 0.412 $\pm$ 0.089 & (5.3 $\pm$ 1.4)E-4 & ...\\
L1448 IRS2 & IRS2 & 50.8 $\pm$ 11.9 & 4.30 $\pm$ 0.66 & 0.483 $\pm$ 0.105 & ... & 2.48 $\pm$ 0.44\\
… & IRS2E & < 30.9 & < 0.12 & < 0.882 & < 4.2E-2 & > 2.67\\
NGC1333 SVS13 & A & 74.6 $\pm$ 51.9 & 119.28 $\pm$ 18.31 & 0.140 $\pm$ 0.030 & (4.2 $\pm$ 0.9)E-4 & ...\\
… & B & 35.5 $\pm$ 0.1 & 10.26 $\pm$ 1.57 & 0.143 $\pm$ 0.031 & (5.1 $\pm$ 1.1)E-3 & ...\\
… & C & 38.0 $\pm$ 3.1 & 2.22 $\pm$ 0.34 & 0.670 $\pm$ 0.145 & (5.8 $\pm$ 1.3)E-3 & > -1.40\\
NGC1333 IRAS2 & IRAS2A & 48.8 $\pm$ 9.5 & 47.06 $\pm$ 7.21 & 0.441 $\pm$ 0.095 & (1.4 $\pm$ 0.3)E-3 & 3.09 $\pm$ 0.54\\
… & IRAS2B & 47.7 $\pm$ 0.5 & 5.27 $\pm$ 0.81 & 0.412 $\pm$ 0.089 & (1.2 $\pm$ 0.3)E-3 & > 2.58\\
NGC1333 IRAS7 & Per18 & 46.4 $\pm$ 12.4 & 4.70 $\pm$ 0.72 & 0.522 $\pm$ 0.114 & (3.6 $\pm$ 0.8)E-3 & 2.20 $\pm$ 0.52\\
… & Per21 & 51.7 $\pm$ 16.5 & 3.42 $\pm$ 0.53 & 0.478 $\pm$ 0.104 & (3.3 $\pm$ 0.7)E-3 & 2.01 $\pm$ 0.36\\
… & Per49 & 315.1 $\pm$ 36.4 & 0.66 $\pm$ 0.11 & 0.058 $\pm$ 0.013 & ... & 0.19 $\pm$ 0.39\\
NGC1333 IRAS4 & B & 34.2 $\pm$ 18.9 & 4.46 $\pm$ 0.68 & 0.778 $\pm$ 0.169 & (1.2 $\pm$ 0.3)E-2 & 1.65 $\pm$ 0.93\\
… & B' & 8.6 $\pm$ 0.1 & 1.76 $\pm$ 0.28 & 0.010 $\pm$ 0.002 & (9.6 $\pm$ 2.1)E-3 & ...\\
NGC1333 IRAS5 & Per52 & 306.8 $\pm$ 88.4 & 0.13 $\pm$ 0.02 & 0.158 $\pm$ 0.036 & ... & 0.34 $\pm$ 0.09\\
… & Per63 & 476.5 $\pm$ 60.8 & 1.52 $\pm$ 0.25 & 0.061 $\pm$ 0.014 & ... & 0.15 $\pm$ 0.17\\
NGC1333 Per6+Per10 & Per6 & 72.7 $\pm$ 12.6 & 0.18 $\pm$ 0.03 & 0.307 $\pm$ 0.068 & ... & 2.18 $\pm$ 0.34\\
… & Per10 & 44.5 $\pm$ 10.0 & 0.44 $\pm$ 0.07 & 0.576 $\pm$ 0.125 & ... & 1.63 $\pm$ 0.08\\
IC348 Per8+Per55 & Per8 & 51.8 $\pm$ 53.7 & 1.86 $\pm$ 0.29 & 0.507 $\pm$ 0.110 & ... & 1.07 $\pm$ 0.23\\
… & Per55 & 334.1 $\pm$ 39.6 & 1.49 $\pm$ 0.25 & 0.068 $\pm$ 0.015 & ... & 0.22 $\pm$ 0.07\\
IC348 MMS & MMS2 & 34.2 $\pm$ 35.7 & 2.23 $\pm$ 0.34 & 0.323 $\pm$ 0.070 & ... & 1.61 $\pm$ 0.33\\
… & E & 35.8 $\pm$ 64.9 & 0.10 $\pm$ 0.03 & 0.734 $\pm$ 0.306 & (8.1 $\pm$ 3.8)E-2 & > 2.34\\
IC348 SMM2 & S & 42.3 $\pm$ 18.8 & 0.93 $\pm$ 0.14 & 0.182 $\pm$ 0.039 & ... & 0.03 $\pm$ 0.68\\
… & N & 47.4 $\pm$ 16.7 & 0.34 $\pm$ 0.05 & 0.584 $\pm$ 0.127 & ... & 1.53 $\pm$ 0.19\\
B1-b & N & 22.0 $\pm$ 0.1 & 0.16 $\pm$ 0.02 & 0.817 $\pm$ 0.177 & (2.0 $\pm$ 0.4)E-1 & > -1.40\\
… & S & 23.5 $\pm$ 11.0 & 0.32 $\pm$ 0.05 & 0.985 $\pm$ 0.216 & (1.2 $\pm$ 0.3)E-1 & > 0.46\\
… & W & 222.3 $\pm$ 16.7 & 0.10 $\pm$ 0.02 & 0.049 $\pm$ 0.012 & ... & 0.70 $\pm$ 0.33\\
NGC1333 Per58+Per65 & Per58 & 278.2 $\pm$ 43.0 & 0.66 $\pm$ 0.11 & 0.121 $\pm$ 0.027 & ... & 0.50 $\pm$ 0.25\\
… & Per65 & 550.6 $\pm$ 58.6 & 0.11 $\pm$ 0.02 & 0.186 $\pm$ 0.042 & ... & -0.06 $\pm$ 0.03\\
IC348Per32+EDJ2009-366 & Per32 & 124.4 $\pm$ 22.4 & 0.06 $\pm$ 0.01 & 0.352 $\pm$ 0.083 & ... & 1.54 $\pm$ 0.14\\
… & EDJ2009-366 & 777.8 $\pm$ 52.8 & 1.23 $\pm$ 0.20 & 0.011 $\pm$ 0.002 & ... & -0.54 $\pm$ 0.09\\
NGC1333 Per37 & Per37 & 36.6 $\pm$ 28.9 & 0.48 $\pm$ 0.07 & 0.687 $\pm$ 0.150 & (2.1 $\pm$ 0.5)E-2 & 1.25 $\pm$ 0.31\\
… & EDJ2009-235 & 291.3 $\pm$ 14.4 & 0.02 $\pm$ 0.00 & ... & ... & 1.02 $\pm$ 0.31\\
… & EDJ2009-233 & 1276.3 $\pm$ 67.3 & 1.33 $\pm$ 0.21 & 0.010 $\pm$ 0.002 & ... & -0.77 $\pm$ 0.04\\
\hline
\end{tabular}
\end{table*}

\subsection{Source properties}
\label{sub:physics}
Source properties derived from the constructed SED are expected to aid in the evolutionary stage classification.
Constraining the peak of an SED improves the calculation of the protostellar system's derived properties, which makes the \textit{Herschel} PACS observations crucial for this task.
Five parameters were derived for each constructed SED: infrared spectral index, bolometric temperature and luminosity and two luminosity ratios.

For the infrared spectral index $\alpha_{\rm IR}$, the slope between 3 $\mu$m and 24 $\mu$m is given by
\begin{displaymath}
\alpha_{\rm IR} = \frac{d~\rm log(\lambda~F_{\lambda})}{d~ \rm log~\lambda}
\end{displaymath}
where $F_{\lambda}$ is the flux at a given wavelength $\lambda$. If the flux at 24 $\mu$m is absent, $\alpha_{\rm IR}$ is not reported. When one or more of the fluxes in this range is an upper limit, $\alpha_{\rm IR}$ is a lower limit.

The bolometric temperature $T_{\rm bol}$ is expressed as
\begin{displaymath}
T_{\rm bol} = 1.25~\times~10^{-11}~\frac{\int_{0}^{\nu}~\nu~S_{\nu}~d\nu}{\int_{0}^{\nu}~S_{\nu}~d\nu}
\end{displaymath}
where $S_{\nu}$ is the flux at a given frequency $\nu$. 

The bolometric luminosity $L_{\rm bol}$ is derived using
\begin{displaymath}
L_{\rm bol} = 4~\pi~D^2~\int_{0}^{\nu}~S_{\nu}~d\nu
\end{displaymath}
where $D$ is the distance. Submillimeter luminosity ($\lambda \geq$ 350 $\mu$m) $L_{\rm submm}$ and far-infrared luminosity ($\lambda \leq$ 70 $\mu$m) $L_{\rm fir}$ were derived from the same equation using the corresponding wavelength ranges.
Both $T_{\rm bol}$ and $L_{\rm bol}$ were derived from the SEDs using trapezoidal integration.

In addition, two luminosity ratios were taken: submillimeter to bolometric $L_{\rm submm} / L_{\rm bol}$ and far-infrared to bolometric $L_{\rm fir} / L_{\rm bol}$.
Both ratios were used since interferometric continuum observations resolve out much of the extended flux pertaining to the envelope, while the far-infrared fluxes from \textit{Herschel} are expected
to capture most of the envelope emission.
These ratios are meant to reflect the envelope to central star mass ratio \citep{andre1993,froebrich2005}, which is used to define the separate physical stages of protostars \citep{robitaille2006,enoch2009}.
Deeply embedded sources are expected to have luminosity ratios higher than 0.005, while less embedded protostars tend to show ratios lower than 0.005.

\subsection{Caveats}
\label{sub:caveats}
The results in this work are limited by the resolution of the \textit{Herschel} PACS maps. Multiple systems with separations $<$7$\arcsec$ cannot be resolved, making the frequency of non-coevality found in this work applicable to wider systems. Furthermore, the results obtained here can provide constraints for multiple protostellar system formation scenarios at the core scale ($\geq$1600 AU). 

Some multiple systems lack reported resolved submillimeter fluxes, which means that the derived properties are under- or overestimated. This affects the evolutionary classification derived from these parameters. Care must then be taken to consider this aspect when classifying the systems, and the relations between components of a system are more relevant than the actual quantities themselves.

\section{Results and analysis}
\label{sec:analysis}
The constructed SEDs are presented in Fig.~\ref{fig:SEDresolved} for resolved systems and in Fig. \ref{fig:SEDunresolved} for unresolved systems. 
Flux uncertainties are in general comparatively small, hence when plotted, the errors are not much larger than the symbols used for plotting.

The constructed SEDs are analyzed with the aim to study the coevality of multiple systems.
All the parameters typically used to identify a protostellar system's evolutionary stage together with additional diagnostics are used in the classification.
This is to ensure that there is as little bias as possible due to inclination, which tends to affect the derived SED parameters.
In this section each method and the corresponding results are presented.

\subsection{SED shapes}
\label{sub:shape}
As the protostellar system evolves and clears out the envelope, the peak of the energy distribution shifts to shorter wavelengths. 
The wavelength at which the SED peaks can therefore be used as an indicator of the evolutionary stage.
Average SEDs for the progressive evolutionary stages are shown in Figure~\ref{fig:avgsed}. 
These SEDs were derived from the \textit{Spitzer} c2d observations of a large sample of protostars in Perseus and Serpens by \citet{enoch2009} and divided into classes based on $T_{\rm bol}$.
Figures~\ref{fig:SEDresolved} and ~\ref{fig:SEDunresolved} show the constructed SEDs for multiples compared with these average SEDs.

A quick look at the constructed SEDs makes it clear that several multiple systems have components with different SED shapes (e.g., IC348 Per8+Per55, IC348 Per32+EDJ2009-366), while others have components with similar SED shapes (e.g., NGC1333 IRAS 5, IC348 MMS2) or a combination (e.g., NGC1333 IRAS 7, L1448 N, NGC1333 SVS13, B1-b).
The similar SEDs hint at coeval components, whereas non-coevality is suggested by the differing SEDs.

To obtain some simple statistics, we counted the systems and identified stages by eye in comparison to each other and to Fig.~\ref{fig:avgsed}.
The frequency of non-coevality found in this way is listed in Table~\ref{tab:SEDstats}.
Higher order multiples were counted twice, once for the first pair and then a second time for the pair compared to the third component.
For example, NGC1333 IRAS7 was counted once as coeval and once as non-coeval, since Per18 and Per21 appear to have the same evolutionary stage, but are non-coeval relative to Per49.
This generates a total of 21 systems where coevality is probed, in contrast to the 16 systems in our sample.
We found that 7 of 21 systems (33 $\pm$ 10\%) show non-ceovality: L1448 N, NGC1333 SVS13, NGC1333 IRAS7, IC348 Per8+Per55, B1-b, IC348 Per32+EDJ2009-366 and NGC1333 Per37.
We did not set a maximum separation limit for a multiple system, but it is interesting to see the change in non-coevality frequency in our sample as a limit is set. Assuming the characteristic size of protostellar cores (30$\arcsec$), we found that 6 of 15 systems (40 $\pm$ 13\%) with separations $\leq$30$\arcsec$ are non-coeval. An arbitrary separation limit of $\leq$20$\arcsec$ shows 4 out of 14 systems (33 $\pm$ 14\%) to be non-coeval. This
means that the rate of non-coevality does not change significantly by limiting the separation of multiple systems.

For NGC1333 Per37, the EDJ2009-235 component is not detected in the \textit{Herschel} PACS maps, but is detected in the c2d and VANDAM surveys. \citet{tobin2016} classified it as a Class II source, but based on the c2d fluxes, EDJ2009-235 appears to be an embedded source, closer in agreement with the classification from \citet{young2015}. A possible explanation for the discrepancy and its lack of \textit{Herschel} PACS detection could be a highly extincted disk that might make a Class II source look much younger.
The IC348 systems Per8+Per55 and Per32+EDJ2009-366, which appear as proto-binaries at scales $\geq$7$\arcsec$, have unresolved components (Table~\ref{tab:unresolved}), which means they are higher order multiples.
This would suggest that higher order multiples tend toward non-coevality.

\begin{figure*}
\centering
\includegraphics[width=\textwidth]{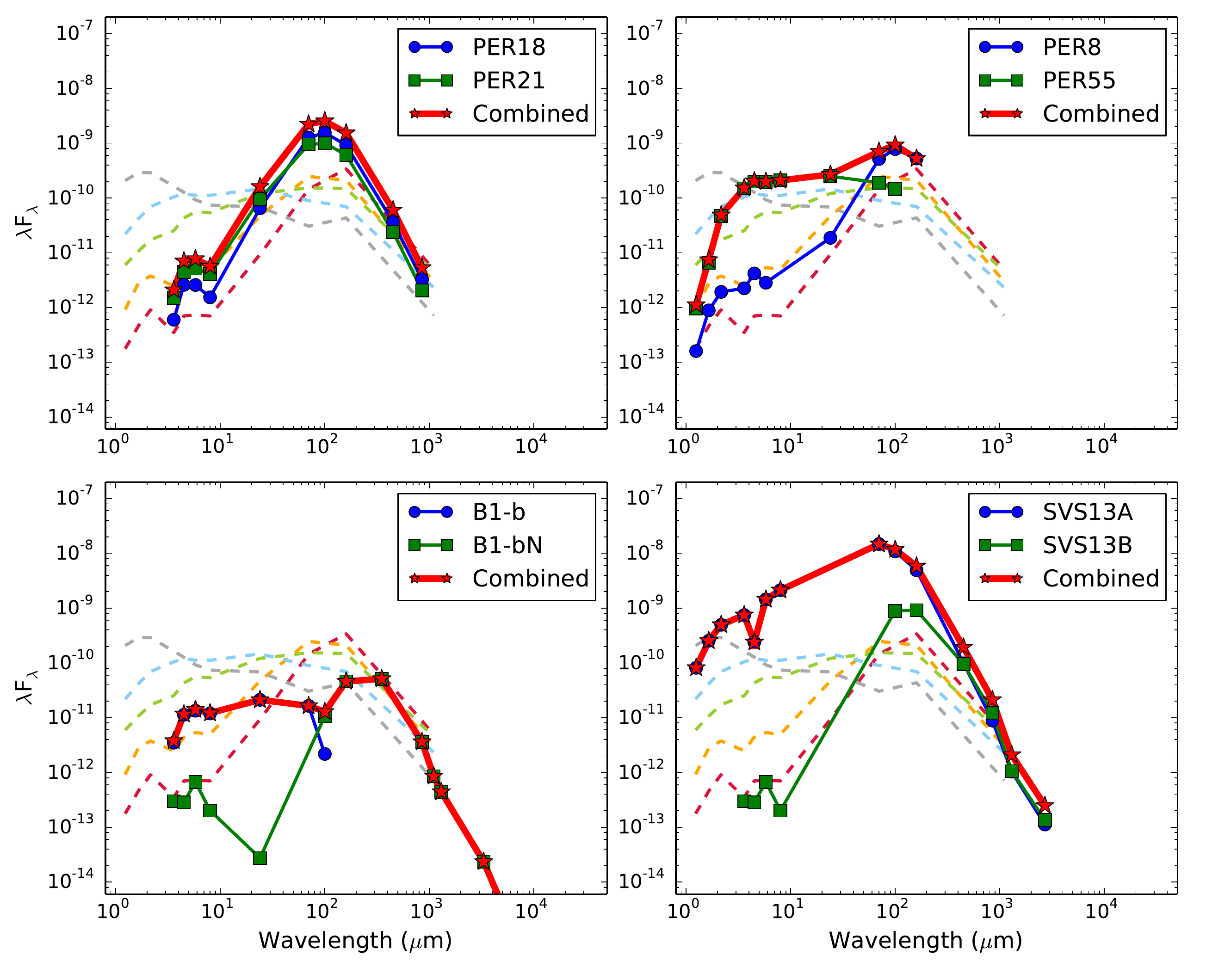}
\caption{Summed SEDs from resolved multiple systems, showing how unresolved multiple systems might be composed. The dashed lines are the same as in Fig.~\ref{fig:avgsed}.}
\label{fig:combSEDs}
\end{figure*}

\begin{table*}
\caption{Derived parameters of combined SEDs}
\label{tab:combSEDs}
\centering
\begin{tabular}{c c c c c c}
\hline \hline
Source & $T_{\rm bol}$ (K) & Class by $T_{\rm bol}$ & $L_{\rm bol}$ (L$_{\odot}$) & $L_{\rm fir}$ / $L_{\rm bol}$ & Class by L-ratio\\
\hline 
IRAS7 Per18 & 46.4 $\pm$ 12.4 & Early 0 & 4.70 $\pm$ 0.72 & 0.52 $\pm$ 0.11 & 0 \\
IRAS7 Per21 & 51.7 $\pm$ 16.5 & Late 0 & 3.42 $\pm$ 0.53 & 0.48 $\pm$ 0.10 & 0 \\
\textbf{IRAS7 Combined} & \textbf{48.7 $\pm$ 11.57} & \textbf{Early 0} & \textbf{8.13 $\pm$ 1.25} & \textbf{0.50 $\pm$ 0.11} & \textbf{0 }\\
\hline
IC348 Per8 & 51.8 $\pm$ 53.7 & Late 0 & 1.86 $\pm$ 0.29 & 0.51 $\pm$ 0.11 & 0 \\
IC348 Per55 & 334.1 $\pm$ 39.6 & Late I & 1.49 $\pm$ 0.25 & 0.068 $\pm$ 0.015 & 0 \\
\textbf{IC348 Combined} & \textbf{175.5 $\pm$ 36.8} & \textbf{Early I} & \textbf{3.38 $\pm$ 0.53} & \textbf{0.32 $\pm$ 0.07} & \textbf{0} \\
\hline 
B1-b & 222.3 $\pm$ 16.7 & Early I & 0.10 $\pm$ 0.017 & 0.049 $\pm$ 0.012 & 0 \\
B1-bN & 22.0 $\pm$ 0.1 & Early 0 & 0.16 $\pm$ 0.02 & 0.82 $\pm$ 0.18 & 0 \\
\textbf{B1-b Combined} & \textbf{106.0 $\pm$ 16.6} & \textbf{Early I} & \textbf{0.24 $\pm$ 0.04} & \textbf{0.59 $\pm$ 0.13} & \textbf{0} \\
\hline
SVS13A & 74.6 $\pm$ 51.9 & Late 0 & 119.28 $\pm$ 18.31 & 0.14 $\pm$ 0.03 & 0 \\
SVS13B & 35.5 $\pm$ 0.1 & Early 0 & 10.26 $\pm$ 1.57 & 0.14 $\pm$ 0.03 & 0 \\
\textbf{SVS13 Combined} & \textbf{73.9 $\pm$ 51.9} & \textbf{Late 0} & \textbf{121.07 $\pm$ 18.58} & \textbf{0.15 $\pm$ 0.03} & \textbf{0} \\
\hline
\end{tabular}
\end{table*}

\subsection{Classification from derived properties}
\label{sub:physclass}
All the calculated parameters described in Sect.~\ref{sub:physics} and their errors are listed for each source in Table~\ref{tab:param} for the resolved multiple systems.
Comparing $T_{\rm bol}$ within the multiple systems indicates that the rate of non-coevality is much lower than found from the visual comparison of the SEDs. 
Based on $T_{\rm bol}$, 6 multiple systems are found to be non-coeval (NGC1333 IRAS7, IC348 Per8+Per55, B1-b, Per32+EDJ2009-366 and NGC1333 Per37 twice). Marginal non-coevality, that is, one source being slightly younger than the other (e.g., early Class 0 and late Class 0), can be seen toward 4 systems (L1448 C, NGC1333 SVS13, NGC1333 Per6+Per10 and NGC1333 Per58+Per65), while the remaining are quite coeval.
The non-coevality frequency found based on $T_{\rm bol}$ is between 29 $\pm$ 10\% and 48 $\pm$ 11\%, with the latter value considering the marginally non-coeval systems in addition to the non-coeval systems.

Luminosity ratios, such as $L_{\rm fir} / L_{\rm bol}$ and $L_{\rm submm} / L_{\rm bol}$, are expected to be indicators of the evolutionary stage, with values above 0.005 indicating a Class 0 source. 
However, for these ratios to provide reliable information, a well-sampled SED at $\lambda$ $\geq$ 70$\mu$m is required.
Not all the systems we studied here are well sampled in this regime. For most systems the $L_{\rm submm} / L_{\rm bol}$ ratio is underestimated or cannot be calculated at all, while the $L_{\rm fir} / L_{\rm bol}$ can be calculated in all cases but is also underestimated due to the lack of submillimeter flux.
Systems relatively well sampled in the submillimeter regime in our sample are NGC1333 SVS13 and B1-bN \& S.
The $L_{\rm submm} / L_{\rm bol}$ ratio for NGC1333 SVS13 supports the non-coevality of this system, with the ratio for SVS13A being lower than 0.5\%, the threshold for embedded sources, while SVS13B and C are (marginally) above the value, indicating they are embedded. 
In contrast, the $L_{\rm fir} / L_{\rm bol}$ ratio shows values well above 0.5\% for all three sources, with SVS13A and B having equal values and SVS13C showing a higher value. 
This would seem to suggest that SVS13C is more embedded than its northern companions, and SVS13A and B are less embedded.
B1-bN \& S have a $L_{\rm submm} / L_{\rm bol}$ well above 0.5\% and a high value for the $L_{\rm fir} / L_{\rm bol}$ ratio, consistent with their deeply embedded condition.
A non-coevality frequency cannot be derived from the luminosity ratios in this work because they are not well determined for almost all sources.

The infrared spectral index $\alpha_{\rm IR}$, which is the slope of the SED between 3 and 24 $\mu$m, is assumed to be a good indicator of evolutionary stage, even when geometric effects affect the SED shape \citep{crapsi2008}. 
Positive values indicate an embedded source (Class 0 and I), while negative values higher and lower than -1.5 point toward Class II and III, respectively.
NGC1333 SVS13A and SVS13B lack a point at 24 $\mu$m and therefore the value was not calculated. SVS13C has a negative value, suggesting that it is a Class II source.
For B1-bN \& S, the northern source presents a negative value higher than -1.5 while the southern source has a positive value, which would indicate that the northern component is Class II and the southern component is an embedded source. 
The reason for these discrepancies relative to the other parameters is the sensitivity of the spectral index to the fluxes between 3 and 24 $\mu$m. 
For example, NGC1333 SVS13C's flux at 24 $\mu$m drops below the fluxes at shorter wavelengths, causing the slope and consequently the spectral index to be negative.
The same occurs for B1-bN , whose flux at 24 $\mu$m is lower than for B1-bS.
Interestingly, this parameter causes systems such as NGC1333 IRAS7 and IC348 Per8+Per55 to seem coeval.
A coevality frequency cannot be derived from the values of $\alpha_{\rm IR}$ for the sample in this work, since it is not well determined for all sources.

The derived properties suggest the non-coevality frequency to be lower than that obtained from a visual examination of the SEDs, when considering only non-coeval systems and not marginally non-coeval ones.
Still, the derived properties are sensitive to inclination and SED sampling, which biases their classification of evolutionary stage.
Hence, these parameters should not be taken at face value, but instead be considered together with the SED shape and other properties of the multiple systems.

\subsection{Resolved versus unresolved SEDs of multiples}
\label{sub:combseds}
The SED of an unresolved multiple system is composed of the sum of SEDs of its individual components. 
This led us to analyze to which extent an unresolved SED reflects the parameters and evolutionary stage classification of its compoennts.
To do so, the SEDs of resolved systems were summed and the resulting shape and derived properties compared to those of the individual components.
This analysis does not generate a method to separate unresolved SEDs, but will provide insight into the coevality of close multiple systems.
Figure~\ref{fig:combSEDs} and Table~\ref{tab:combSEDs} show the results of combining the SEDs of four systems: NGC1333 IRAS7 Per18 and Per21, IC348 Per55 and Per8, B1-bW and B1-bN, and NGC1333 SVS13 A and B.

From this simple analysis we find that there are mainly three cases. 
The first is that if the two components have almost identical SEDs, then the combined SED will simply be doubled. 
This case is shown by NGC1333 IRAS7 Per18 and Per21. 
The second case is when the two components are non-coeval, then the SED will not follow a specific SED shape but will appear odd shaped with two peaks. 
This is illustrated by the combined SEDs of IC348 Per8 and Per55, and B1-b and B1-bN. 
The final case occurs when one component is noticeably dimmer and younger than the other, then the brightest component dominates the combined SED. NGC1333 SVS13 A and B illustrate this scenario. 
Thus, different components can dominate different regions of the SED.

Figure~\ref{fig:SEDunresolved} presents the SEDs of unresolved multiple systems.
Examples of the second case from Fig.~\ref{fig:combSEDs}, such as L1455 FIR2, B1-a and Per62, are shown in Fig.~\ref{fig:SEDunresolved}. 
The first and third cases are next to impossible to identify without additional constraints, and systems such as IRAS03292, IC348 SMM22, EDJ2009-156, and HH211 could be examples of either case.

In all scenarios we find that $L_{\rm bol}$ for the unresolved SED is equal to the sum of both components. 
In contrast, parameters such as $T_{\rm bol}$ and $L_{\rm fir}$ / $L_{\rm bol}$ are an arithmetic average of the corresponding parameters of the two components. 
This, of course, affects the evolutionary classification of unresolved multiple systems.
While taking the derived values and assuming each component contributes equally may be a good assumption in
coeval cases, this could be an over- or underestimation of the true parameters in non-coeval systems.

\section{Discussion}
\label{sec:discuss}
From the SED shapes of resolved multiple systems alone, we find a non-coevality frequency of 33 $\pm$ 10\%. 
Higher order multiple systems contribute the most to this fraction, with all of five resolved systems (L1448 N, NGC1333 SVS13, NGC1333 IRAS7, B1-b and NGC1333 Per37) showing indications of non-coevality.
The other two systems, IC348 Per8+Per55 and Per32+EDJ2009-366, that appear as binaries at separations $\geq$ 7$\arcsec$, show additional fragmentation at scales $<$7$\arcsec$ in one of the components (Table~\ref{tab:unresolved}), which also makes them higher order multiples.
Binaries, on the other hand, tend toward coevality in our sample.
The non-coevality frequency found here is similar to that found by \citet{kraus2009} of one-third in pre-main sequence stars.
However, their frequency was found from binaries alone, whereas higher order multiple systems are responsible for this frequency in our study.
\citet{kraus2009} also probed down to small separations ($>$200 AU) an order of magnitude lower than the separations probed here ($\geq$1600 AU). 
The unresolved systems studied here that are suspected of non-coevality could account for the rest of that fraction.
The question then arises whether the non-coevality frequency obtained here is real or a product of misalignment, that is, the difference in inclination (w.r.t. the line of sight) of each component in a multiple system.

Different SEDs do not necessarily indicate a non-coeval multiple system, but could also be due to geometrical effects, especially if the line of sight lies through the outflow cone.
Inclination can alter the shape of the SED and derived parameters \citep{robitaille2006,crapsi2008,enoch2009} and is crucial to attain an accurate determination of the evolutionary stage \citep{offner2012b}.
If the protostellar system is seen edge-on, the protostar is obscured and will seem younger.
On the other hand, if the protostellar system is observed face-on, the protostar will be unobscured and appear more evolved.
Hence, the alignment of multiple systems affects whether the differing SED shapes are a product of real non-coevality of inclination effects.
Multiple system formation mechanisms suggest that ordered rotational fragmentation would produce systems with aligned inclination \citep{burkert1993}, whereas turbulent fragmentation is expected to produce random alignment \citep{offner2010}.
Work on pre-main sequence multiple systems shows that binaries have a tendency to be aligned, while higher order multiple systems are less likely to be aligned \citep{jensen2004,monin2006}.
However, the numbers seen toward pre-main sequence multiple systems might not reflect the actual alignment at the time of formation
because of the dynamical evolution \citep{reipurth2000,jensen2004}.
Although a handful of observations exist that show both aligned and misaligned multiple systems at every stage of evolution, there are no statistical numbers on the distribution of inclination and multiple system alignment at the time of formation.

The best method to obtain an accurate inclination estimate is through disk observations and modeling, but not all protostellar systems have confirmed and reported disks.
Hence, the inclination of protostellar systems must be constrained through another technique.

\subsection{Outflows}
\label{sub:outflows}
Outflows present a viable option, but they can provide only a broad inclination range and may not always be accurate, even more so in multiple systems where precession occurs due to companion perturbations \citep{fendt1998}.
The evolutionary stage of a protostar is closely linked to its outflow and circumstellar envelope.
The envelope is dispersed as the protostar evolves and accretes part of its material \citep{andre1993}, while the molecular outflow tends to become weaker with time and the outflow cavity broadens \citep{velusamy1998,arce2006}.

A point to highlight here is that misalignment of outflow axes on the plane of the sky is not the same as misalingment of inclination with respect to the line of sight. The former does not affect the observed SED, while the latter has a strong effect on a protostar's SED and derived parameters. Thus, we refer to the first case as the outflow position angle (P.A.) and the second case as inclination misalignment.

\citet{lee2016} studied the outflows of 9 multiple systems in Perseus and found in all cases that the outflows of wide multiple systems have different P.A., that is, that they are misaligned on the plane of the sky. However, strong indications of inclination misalignment were not found.

To compare with the other parameters used to determine the evolutionary stage of components in a system, we explicitly examine here the outflows of a few systems, focusing on signs that are expected to indicate evolutionary stage and inclination misalignment. 
Specific comments on each resolved multiple system treated in this work are given in Appendix~\ref{app:systems}.

NGC1333 SVS13A shows molecular outflow lobes that are wide and shell-like, while SVS13C exhibits a collimated outflow with indications of being in the plane of the sky, meaning that the disk is seen
edge-on \citep{plunkett2013,lee2016}.
This possible inclination misalignment might be the reason that this system appears to be non-coeval based on the SEDs. 
However, SVS13A has been suggested to be a transition class0/I object based on its association with a Haro-Herbig object and the shell-like morphology of its outflow, making it somewhat older than SVS13B and C.

B1-bN and S have outflows that appear parallel to each other, but the blueshifted lobes are in opposite directions, a tell-tale sign of inclination misalignment.
Even though their outflows suggest inclination misaglinment \citep{gerin2015}, the SEDs appear to be similar, indicating coevality, which is consistent with the results obtained from the analysis of their environment \citep{hirano2014}. The inclination misalignment may therefore be small.
B1-bW is expected to be older based on the SED, which is why
it is not detected in the submillimeter because it has too little envelope material that might be the product of stripping from a neighboring outflow or jet \citep{hirano2014}.

This shows that inclination misalignment of the systems does not always generate an apparent non-coevality. To do this, one of the components of a system would need to be significantly inclined, tending toward the line of sight along the outflow cavity so that the source would appear much older.

\subsection{Alignment and coevality}
\label{sub:alignment}
To assess the frequency of apparent non-coevality due to misalignment versus real non-coevality, we made a simple statistical test with the \citet{robitaille2006} SED model grid\footnote{Retrieved Oct. 2015 from http://caravan.astro.wisc.edu/protostars/}.
This grid of models offers 10 inclination angles, evenly sampled in $cos(i)$, ranging from 87.13$^{\circ}$ to 18.19$^{\circ}$ with 0$^{\circ}$ being the disk seen face-on (i.e., looking down the outflow cone), which is ideally suited for our task.
We constrained the number of models by choosing those for Class 0 sources, filtering by 2.0 M$_{\odot}$ $\leq M_{\rm envelope} \leq$ 10.0 M$_{\odot}$ and $M_{\rm envelope} > M_{\rm star}$.
The resulting set of model SEDs, a total of 1037 times 10 inclination angles, have stellar masses of up to 2.0 M$_{\odot}$ and outflow cavity angles ranging from 15 to 30$^{\circ}$.
The fluxes are obtained for similar wavelengths as our observations, including the \textit{Herschel} PACS fluxes, and an aperture of 7000 AU (30$\arcsec$ at $d \sim$ 235 pc).

\begin{figure*}
\includegraphics[width=\textwidth]{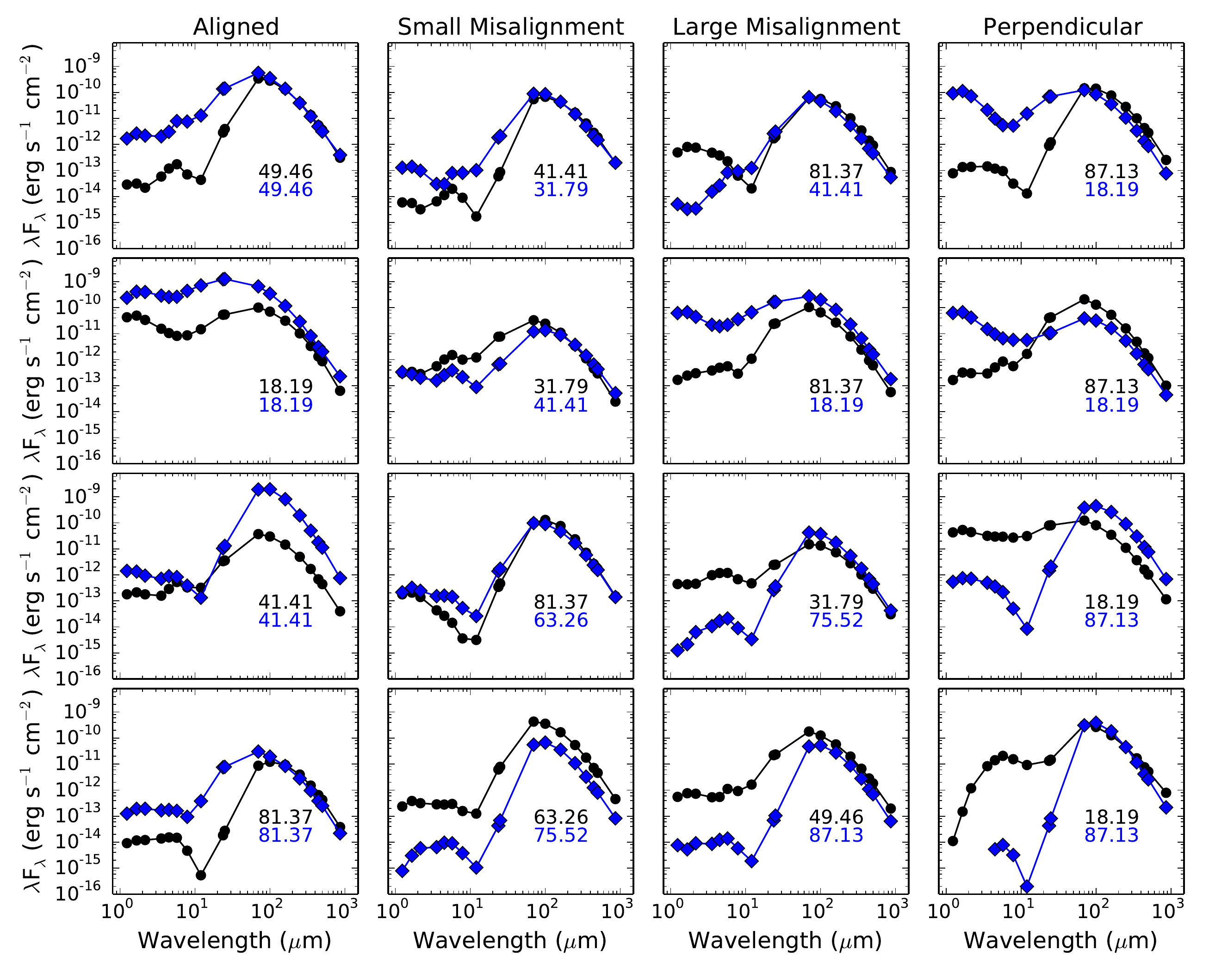}
\caption{Randomly paired model SEDs in the four groups based on alignment. The numbers indicate the inclination of the respective component. Note that the SED pairs that appear non-coeval peak at the same wavelength for both components, which is not expected from components at different evolutionary stages.}
\label{fig:modSEDs}
\end{figure*}

Pairs of models with their respective inclinations are randomly drawn from the list of 10370 Class 0 synthetic SEDs, resulting in 5185 pairs.
This is done to simulate, in a simplified manner, multiple systems and compare their SEDs and derived parameters.
The random paired models are separated into four groups determined by the difference in inclination angles, that is, their degree of alignment.
The four groups are perfectly aligned ($\Delta i$ = 0$^{\circ}$), small misalignment (0$^{\circ}$ < $\Delta i$ < 34$^{\circ}$), large misalignment (34$^{\circ}$ < $\Delta i$ < 69$^{\circ}$) and perpendicular ($\Delta i$ = 69$^{\circ}$).
Perpendicular alignment is not equal to 90$^{\circ}$ because
of the available inclination angles of the models, but is instead the difference between the edge-on (87.13$^{\circ}$) and face-on cases (18.19$^{\circ}$).
To obtain the frequency of pairs in each group, the times each case occurs were counted based on $\Delta i$ and are listed in Table~\ref{tab:stats}.
Perpendicular alignment is the least likely case (2 $\pm$ 0.2\%), with small alignment the most common (63 $\pm$ 0.6\%).
Examples of SEDs from each case are shown in Fig.~\ref{fig:modSEDs}.

Apparent non-coevality was checked by first filtering with $T_{\rm bol}$, assuming that for apparently non-coeval pairs the $T_{\rm bol}$ difference is larger than a factor of 3.
The reason for using $T_{\rm bol}$ to filter the models is based on three points: i) $T_{\rm bol}$ tends to be sensitive to inclination, ii) the thresholds for evolutionary stage classification (late Class 0 to early Class I: 100 K; late Class I to Class II: 650 K), and iii) the non-coeval resolved multiple systems in our sample identified from $T_{\rm bol}$ have a ratio of about 6 or higher.
A factor of 3 was chosen to ensure that Class 0 and I pairs are also included, since for example a component with a $T_{\rm bol}$ = 50 K may appear non-coeval with a companion having a $T_{\rm bol}$ > 150 K.
The paired SEDs filtered this way were then inspected by eye to subtract the SED pairs that did not appear non-coeval.
The frequency of apparent non-coevality due to misalignment is found to be 17\% $\pm$ 0.5\%.
Examining the frequency of apparent non-coevality in each of the four groups, we found that large misalignment and perpendicular have the most common occurrences of apparent non-coevality.
This is mainly due to one component being face-on or close to face-on, in combination with the outflow cavity opening angle, causing one component to appear older.
This was also suggested by the results of the outflows of multiple systems.

A characteristic of SEDs for different evolutionary stages is that the peak of the SED shifts to shorter wavelengths as the envelope is dispersed (see Fig.~\ref{fig:avgsed}).
While the SEDs can appear non-coeval as a result of inclination effects, the inspection by eye of these SED pairs revealed that the peak around $\lambda \sim$100 $\mu$m, characteristic of
Class 0 sources, does not significantly shift to shorter wavelengths as a result of inclination, even in the face-on case.
Examples of this are shown in Fig.~\ref{fig:modSEDs}.
In other words, even though one of the SEDs in the pair appeared more evolved than the companion, this apparently older source retained its peak around $\lambda \sim$100 $\mu$m.
This is in contrast to the SEDs of non-coeval resolved multiple systems discussed in this work. For example, in NGC1333 IRAS7, the peak of Per49 is located around 5.8 $\mu$m, whereas the peaks of Per18 and Per21 are around 100 $\mu$m. The same is true for B1-b, where B1-bW has a peak at 8 $\mu$m, while B1-bN \& S peak at around 100 $\mu$m.
On the other hand, NGC1333 SVS13 and L1448 N might appear non-coeval as a result of misalignment, given that the SED peaks of all three components are at about the same wavelength. 
For NGC1333 SVS13, the outflow and continuum detections of this object suggest that it might be transitioning to the Class I stage and therefore b slightly non-coeval with its companions.

\subsection{On coevality and non-coevality}
The frequency of non-coevality found in the sample of resolved multiple systems studied in this work can be safely assumed to be due to real non-coevality and not solely to misalignment, since most sources are expected to present small misalignment rather than one component close to face-on.
Non-coevality in our resolved sample is exhibited by higher order multiples, except for IC348 MMS2, a triple with a component at a separation of $\sim$3$\arcsec$ toward the western source (Table~\ref{tab:unresolved}) that appears coeval. 
Proto-binaries, on the other hand, tend toward coevality.
Hence, protostellar siblings most of the time form and probably evolve simultaneously.
This presents some interesting constraints to multiple system formation mechanisms and also raises questions.

For a multiple system to be non-coeval, the companion must either be formed after the first source or binary. 
In other words, fragmentation in the core must occur after the initial collapse and formation of the first protostar or protobinary.
Hydrodynamical simulations predict that heated gas reduces the chance of further fragmentation, with only the cold gas tending to fragment \citep{stamatellos2009b,offner2010,bate2012}.
A possible explanation is that gas heating occurs along the outflow cavity while the dense envelope reduces heating of the surrounding gas, allowing further fragmentation to occur.
Turbulence, on the other hand, is thought to be able to produce non-coevality through random density enhancements in the core.

For the coeval systems found in our resolved sample, fragmentation of the core would have occurred during the initial collapse and the system remained stable enough to hinder any further fragmentation and formation of younger companions, either through heated gas or lack of density enhancements and strong enough turbulence.
Observational evidence for gas heating along the outflow cavity walls has been provided by \citet{vankempen2009} and \citet{yildiz2015}, including some of the multiple systems studied here.

Dynamical evolution, that is, the interaction that occurs in multiple systems, can cause these systems to evolve non-coevally, for example by expelling one of the companions, considerably reducing its envelope and thus truncating its accretion of material \citep{reipurth2010,reipurth2012}.
However, for embedded systems, not enough time has elapsed for dynamical evolution to play a major role in the appearance of the system.
External factors, such as neighbouring outflows and jets, can affect a protostar in a system, for example by stripping material or possibly triggering further fragmentation.
While these mechanisms, dynamical evolution and external factors, are not formation mechanisms, they can alter the conditions of a multiple system and cause it to evolve non-coevally.

Given that only about a third of the multiple systems present non-coevality, the question then arises which factor or factors contribute to making some regions fragment and collapse even
more while others do not. Probing the distribution of heated gas around multiple and single protostellar systems, in the case of multiples for both coeval and non-coeval, could provide insight.

\begin{table}
\caption{Random pair statistics based on synthetic SEDs}
\label{tab:stats}
\centering
\begin{tabular}{c c c}
\hline \hline 
 & Value & Aparent \\
 &  & Non-coevality \\
\hline
Total number of models & 10370 & ... \\
Total number of pairs & 5185 & ... \\
Aligned ($\Delta$i = 0$^{\circ}$) & 9.5$\pm$0.4\% & 8.4$\pm$1.2\% \\
Small misalignment (0$^{\circ}$ < $\Delta i$ < 34$^{\circ}$) & 63.0$\pm$0.6\% & 12.4$\pm$0.6\% \\
Large misalignment (34$^{\circ}$ < $\Delta i$ < 69$^{\circ}$) & 25.5$\pm$0.6\% & 39.5$\pm$1.3\% \\
Perpendicular ($\Delta i$ = 69$^{\circ}$) & 2.0$\pm$0.2\% & 83.9$\pm$3.6\% \\
Total & ... & 17 $\pm$ 0.5\% \\
\hline
\end{tabular}
\end{table}

\section{Conclusions}
\label{sec:conc}
This work presents the constructed SEDs of known resolved multiple protostellar systems in Perseus with separations $\geq$ 7$\arcsec$. The SEDs were constructed from \textit{Herschel} PACS photometric maps and \textit{Spitzer} c2d catalogs, together with fluxes from the literature for the longer wavelengths. 
The properties were then derived from the observed SEDs.
The SED shape and derived parameters were taken together with literature work on the envelope and outflow of these systems to determine the coevality of multiple systems.
The literature, both from observations and models, lacks statistics on the frequency of alignment of multiple systems, but work is ongoing.
A simple test whether the different SEDs might be due to misalignment was carried out in this work by randomly pairing model Class 0 sources from model SED grids with different inclinations with
respect to the line of sight \citep{robitaille2006} and then counting the frequency of apparently non-coeval systems.
The results of this work can be summarized in the following points.

\begin{enumerate}
\item From our sample of resolved multiple protostellar systems, which have separations $\geq$7$\arcsec$, a coevality frequency of 66 $\pm$ 10\% is found, suggesting that most wide multiples are born together.

\item From the observed SED shapes alone, a non-coevality frequency of 33 $\pm$ 10\% is found, with higher order multiples being responsible for this percentage. Random pairing of model SEDs indicates that the frequency of apparent non-coevality that is
due to misalignment of the components' inclinations is 17 $\pm$ 0.5\%, with most occurrences in systems with large misalignments and perpendicular orientations. But most pairs tend toward small misalignment (63 $\pm$ 0.6\%). This indicates that the observed non-coevality toward multiple systems in our sample is not due to misalignment, but is instead real.

\item Derived properties, such as $T_{\rm bol}$ and $\alpha_{\rm IR}$ , suggest that the non-coevality frequency may be lower (21 $\pm$ 9\%). However, the parameters derived from the observed SEDs produce contradicting results. Physical parameters derived from the synthetic SEDs demonstrate that these parameters are very sensitive to inclination or do not provide clear-cut evolutionary stage separations. As previously found in \citet{offner2012b}, the parameters are therefore unreliable for evolutionary stage classification, unless
the inclination is well constrained.

\item Unresolved SEDs from multiple systems are not always dominated by the primary or brightest component, but can present an odd double-peaked shape that is due to non-coeval components. This can alter the fraction of non-coevality, but we did not take this into account because of the high uncertainty.
\end{enumerate}

Owing to the limit on multiple system separations in this work, our results can only place constraints on formation mechanisms at core scales ($\geq$ 1600 AU). 
Higher order multiples show a stronger tendency to be non-coeval.
This suggests that fragmentation at core scales can occur at different times, thus generating these types of systems.

The main conclusion of this work is then that real non-coeval multiple protostellar systems exist in the early stages of protostellar formation at core scales, which is most of the time due to formation, and can be enhanced by dynamical evolution (e.g., component ejection or envelope stripping by external influence, such as an outflow.). 
Several questions then arise. What causes some cores to fragment and collapse at different times? Which role do the already formed protostars play on the formation of their companions? Does temperature play an important role, and if so, to which extent?
Future work, both from observations and models, is needed to address these questions.

\begin{acknowledgements}
Astrochemistry in Leiden is supported by the Netherlands Research School for Astronomy (NOVA), by a Royal Netherlands Academy of Arts and Sciences (KNAW) professor prize, and by the European Union A-ERC grant 291141 CHEMPLAN. This work is supported by grant 639.041.335 from the Netherlands Organisation for Scientific Research (NWO) and by the Netherlands Research School for Astronomy (NOVA) and by the Space Research Organization Netherlands (SRON).
J.J.T is currently supported by grant 639.041.439 from the Netherlands Organisation for Scientific Research (NWO).
DF acknowledges support from the Italian Ministry of Science and Education (MIUR), project SIR (RBSI14ZRHR).
\end{acknowledgements}

\bibliographystyle{aa}
\bibliography{SEDpaper.bib}

\begin{thebibliography}{81}
\expandafter\ifx\csname natexlab\endcsname\relax\def\natexlab#1{#1}\fi

\bibitem[{{Andr{\'e}} {et~al.}(2010){Andr{\'e}}, {Men'shchikov}, {Bontemps},
  {K{\"o}nyves}, {Motte}, {Schneider}, {Didelon}, {Minier}, {Saraceno},
  {Ward-Thompson}, {di Francesco}, {White}, {Molinari}, {Testi}, {Abergel},
  {Griffin}, {Henning}, {Royer}, {Mer{\'{\i}}n}, {Vavrek}, {Attard},
  {Arzoumanian}, {Wilson}, {Ade}, {Aussel}, {Baluteau}, {Benedettini},
  {Bernard}, {Blommaert}, {Cambr{\'e}sy}, {Cox}, {di Giorgio}, {Hargrave},
  {Hennemann}, {Huang}, {Kirk}, {Krause}, {Launhardt}, {Leeks}, {Le Pennec},
  {Li}, {Martin}, {Maury}, {Olofsson}, {Omont}, {Peretto}, {Pezzuto}, {Prusti},
  {Roussel}, {Russeil}, {Sauvage}, {Sibthorpe}, {Sicilia-Aguilar}, {Spinoglio},
  {Waelkens}, {Woodcraft}, \& {Zavagno}}]{andre2010}
{Andr{\'e}}, P., {Men'shchikov}, A., {Bontemps}, S., {et~al.} 2010, \aap, 518,
  L102

\bibitem[{{Andr\'e} {et~al.}(1993){Andr\'e}, {Ward-Thompson}, \&
  {Barsony}}]{andre1993}
{Andr\'e}, P., {Ward-Thompson}, D., \& {Barsony}, M. 1993, \apj, 406, 122

\bibitem[{{Arce} \& {Sargent}(2006)}]{arce2006}
{Arce}, H.~G. \& {Sargent}, A.~I. 2006, \apj, 646, 1070

\bibitem[{{Audard} {et~al.}(2014){Audard}, {{\'A}brah{\'a}m}, {Dunham},
  {Green}, {Grosso}, {Hamaguchi}, {Kastner}, {K{\'o}sp{\'a}l}, {Lodato},
  {Romanova}, {Skinner}, {Vorobyov}, \& {Zhu}}]{audard2014}
{Audard}, M., {{\'A}brah{\'a}m}, P., {Dunham}, M.~M., {et~al.} 2014, Protostars
  and Planets VI, 387

\bibitem[{{Balog} {et~al.}(2014){Balog}, {M{\"u}ller}, {Nielbock}, {Altieri},
  {Klaas}, {Blommaert}, {Linz}, {Lutz}, {Mo{\'o}r}, {Billot}, {Sauvage}, \&
  {Okumura}}]{balog2014}
{Balog}, Z., {M{\"u}ller}, T., {Nielbock}, M., {et~al.} 2014, Experimental
  Astronomy, 37, 129

\bibitem[{{Barsony} {et~al.}(1998){Barsony}, {Ward-Thompson}, {Andr{\'e}}, \&
  {O'Linger}}]{barsony1998}
{Barsony}, M., {Ward-Thompson}, D., {Andr{\'e}}, P., \& {O'Linger}, J. 1998,
  \apj, 509, 733

\bibitem[{{Bate}(2012)}]{bate2012}
{Bate}, M.~R. 2012, \mnras, 419, 3115

\bibitem[{{Burkert} \& {Bodenheimer}(1993)}]{burkert1993}
{Burkert}, A. \& {Bodenheimer}, P. 1993, \mnras, 264, 798

\bibitem[{{Chen} {et~al.}(2013){Chen}, {Arce}, {Zhang}, {Bourke}, {Launhardt},
  {J{\o}rgensen}, {Lee}, {Foster}, {Dunham}, {Pineda}, \& {Henning}}]{chen2013}
{Chen}, X., {Arce}, H.~G., {Zhang}, Q., {et~al.} 2013, \apj, 768, 110

\bibitem[{{Chen} {et~al.}(2010){Chen}, {Arce}, {Zhang}, {Bourke}, {Launhardt},
  {Schmalzl}, \& {Henning}}]{chen2010}
{Chen}, X., {Arce}, H.~G., {Zhang}, Q., {et~al.} 2010, \apj, 715, 1344

\bibitem[{{Chen} {et~al.}(2009){Chen}, {Launhardt}, \& {Henning}}]{chen2009}
{Chen}, X., {Launhardt}, R., \& {Henning}, T. 2009, \apj, 691, 1729

\bibitem[{{Chini} {et~al.}(2012){Chini}, {Hoffmeister}, {Nasseri}, {Stahl}, \&
  {Zinnecker}}]{chini2012}
{Chini}, R., {Hoffmeister}, V.~H., {Nasseri}, A., {Stahl}, O., \& {Zinnecker},
  H. 2012, \mnras, 424, 1925

\bibitem[{{Ciardi} {et~al.}(2003){Ciardi}, {Telesco}, {Williams}, {Fisher},
  {Packham}, {Pi{\~n}a}, \& {Radomski}}]{ciardi2003}
{Ciardi}, D.~R., {Telesco}, C.~M., {Williams}, J.~P., {et~al.} 2003, \apj, 585,
  392

\bibitem[{{Crapsi} {et~al.}(2008){Crapsi}, {van Dishoeck}, {Hogerheijde},
  {Pontoppidan}, \& {Dullemond}}]{crapsi2008}
{Crapsi}, A., {van Dishoeck}, E.~F., {Hogerheijde}, M.~R., {Pontoppidan},
  K.~M., \& {Dullemond}, C.~P. 2008, \aap, 486, 245

\bibitem[{{Curtis} {et~al.}(2010){Curtis}, {Richer}, {Swift}, \&
  {Williams}}]{curtis2010}
{Curtis}, E.~I., {Richer}, J.~S., {Swift}, J.~J., \& {Williams}, J.~P. 2010,
  \mnras, 408, 1516

\bibitem[{{Decin} {et~al.}(2015){Decin}, {Richards}, {Neufeld}, {Steffen},
  {Melnick}, \& {Lombaert}}]{decin2015}
{Decin}, L., {Richards}, A.~M.~S., {Neufeld}, D., {et~al.} 2015, \aap, 574, A5

\bibitem[{{Diolaiti} {et~al.}(2000){Diolaiti}, {Bendinelli}, {Bonaccini},
  {Close}, {Currie}, \& {Parmeggiani}}]{diolaiti2000}
{Diolaiti}, E., {Bendinelli}, O., {Bonaccini}, D., {et~al.} 2000, in Society of
  Photo-Optical Instrumentation Engineers (SPIE) Conference Series, Vol. 4007,
  Adaptive Optical Systems Technology, ed. P.~L. {Wizinowich}, 879--888

\bibitem[{{Duch{\^e}ne} \& {Kraus}(2013)}]{duchene2013}
{Duch{\^e}ne}, G. \& {Kraus}, A. 2013, \araa, 51, 269

\bibitem[{{Duch{\^e}ne} {et~al.}(1999){Duch{\^e}ne}, {Monin}, {Bouvier}, \&
  {M{\'e}nard}}]{duchene1999}
{Duch{\^e}ne}, G., {Monin}, J.-L., {Bouvier}, J., \& {M{\'e}nard}, F. 1999,
  \aap, 351, 954

\bibitem[{{Dunham} {et~al.}(2015){Dunham}, {Allen}, {Evans},
  {Broekhoven-Fiene}, {Cieza}, {Di Francesco}, {Gutermuth}, {Harvey},
  {Hatchell}, {Heiderman}, {Huard}, {Johnstone}, {Kirk}, {Matthews}, {Miller},
  {Peterson}, \& {Young}}]{dunham2015}
{Dunham}, M.~M., {Allen}, L.~E., {Evans}, II, N.~J., {et~al.} 2015, \apjs, 220,
  11

\bibitem[{{Dunham} {et~al.}(2008){Dunham}, {Crapsi}, {Evans}, {Bourke},
  {Huard}, {Myers}, \& {Kauffmann}}]{dunham2008}
{Dunham}, M.~M., {Crapsi}, A., {Evans}, II, N.~J., {et~al.} 2008, \apjs, 179,
  249

\bibitem[{{Dunham} {et~al.}(2014){Dunham}, {Stutz}, {Allen}, {Evans},
  {Fischer}, {Megeath}, {Myers}, {Offner}, {Poteet}, {Tobin}, \&
  {Vorobyov}}]{dunham2014}
{Dunham}, M.~M., {Stutz}, A.~M., {Allen}, L.~E., {et~al.} 2014, Protostars and
  Planets VI, 195

\bibitem[{{Eisl{\"o}ffel} {et~al.}(2003){Eisl{\"o}ffel}, {Froebrich}, {Stanke},
  \& {McCaughrean}}]{eisloffel2003}
{Eisl{\"o}ffel}, J., {Froebrich}, D., {Stanke}, T., \& {McCaughrean}, M.~J.
  2003, \apj, 595, 259

\bibitem[{{Enoch} {et~al.}(2009){Enoch}, {Evans}, {Sargent}, \&
  {Glenn}}]{enoch2009}
{Enoch}, M.~L., {Evans}, II, N.~J., {Sargent}, A.~I., \& {Glenn}, J. 2009,
  \apj, 692, 973

\bibitem[{{Evans} {et~al.}(2009){Evans}, {Dunham}, {J{\o}rgensen}, {Enoch},
  {Mer{\'{\i}}n}, {van Dishoeck}, {Alcal{\'a}}, {Myers}, {Stapelfeldt},
  {Huard}, {Allen}, {Harvey}, {van Kempen}, {Blake}, {Koerner}, {Mundy},
  {Padgett}, \& {Sargent}}]{evans2009}
{Evans}, II, N.~J., {Dunham}, M.~M., {J{\o}rgensen}, J.~K., {et~al.} 2009,
  \apjs, 181, 321

\bibitem[{{Fendt} \& {Zinnecker}(1998)}]{fendt1998}
{Fendt}, C. \& {Zinnecker}, H. 1998, \aap, 334, 750

\bibitem[{{Froebrich}(2005)}]{froebrich2005}
{Froebrich}, D. 2005, \apjs, 156, 169

\bibitem[{{Gerin} {et~al.}(2015){Gerin}, {Pety}, {Fuente}, {Cernicharo},
  {Commer{\c c}on}, \& {Marcelino}}]{gerin2015}
{Gerin}, M., {Pety}, J., {Fuente}, A., {et~al.} 2015, \aap, 577, L2

\bibitem[{{Hartigan} \& {Kenyon}(2003)}]{hartigan2003}
{Hartigan}, P. \& {Kenyon}, S.~J. 2003, \apj, 583, 334

\bibitem[{{Hiramatsu} {et~al.}(2010){Hiramatsu}, {Hirano}, \&
  {Takakuwa}}]{hiramatsu2010}
{Hiramatsu}, M., {Hirano}, N., \& {Takakuwa}, S. 2010, \apj, 712, 778

\bibitem[{{Hirano} {et~al.}(2010){Hirano}, {Ho}, {Liu}, {Shang}, {Lee}, \&
  {Bourke}}]{hirano2010}
{Hirano}, N., {Ho}, P.~P.~T., {Liu}, S.-Y., {et~al.} 2010, \apj, 717, 58

\bibitem[{{Hirano} \& {Liu}(2014)}]{hirano2014}
{Hirano}, N. \& {Liu}, F.-c. 2014, \apj, 789, 50

\bibitem[{{Hirota} {et~al.}(2008){Hirota}, {Bushimata}, {Choi}, {Honma},
  {Imai}, {Iwadate}, {Jike}, {Kameya}, {Kamohara}, {Kan-Ya}, {Kawaguchi},
  {Kijima}, {Kobayashi}, {Kuji}, {Kurayama}, {Manabe}, {Miyaji}, {Nagayama},
  {Nakagawa}, {Oh}, {Omodaka}, {Oyama}, {Sakai}, {Sasao}, {Sato}, {Shibata},
  {Tamura}, \& {Yamashita}}]{hirota2008}
{Hirota}, T., {Bushimata}, T., {Choi}, Y.~K., {et~al.} 2008, \pasj, 60, 37

\bibitem[{{Hirota} {et~al.}(2011){Hirota}, {Honma}, {Imai}, {Sunada}, {Ueno},
  {Kobayashi}, \& {Kawaguchi}}]{hirota2011}
{Hirota}, T., {Honma}, M., {Imai}, H., {et~al.} 2011, \pasj, 63, 1

\bibitem[{{Hsieh} {et~al.}(2015){Hsieh}, {Lai}, {Belloche}, {Wyrowski}, \&
  {Hung}}]{hsieh2015}
{Hsieh}, T.-H., {Lai}, S.-P., {Belloche}, A., {Wyrowski}, F., \& {Hung}, C.-L.
  2015, \apj, 802, 126

\bibitem[{{Hull} {et~al.}(2014){Hull}, {Plambeck}, {Kwon}, {Bower},
  {Carpenter}, {Crutcher}, {Fiege}, {Franzmann}, {Hakobian}, {Heiles}, {Houde},
  {Hughes}, {Lamb}, {Looney}, {Marrone}, {Matthews}, {Pillai}, {Pound},
  {Rahman}, {Sandell}, {Stephens}, {Tobin}, {Vaillancourt}, {Volgenau}, \&
  {Wright}}]{hull2014}
{Hull}, C.~L.~H., {Plambeck}, R.~L., {Kwon}, W., {et~al.} 2014, \apjs, 213, 13

\bibitem[{{Jensen} {et~al.}(2004){Jensen}, {Mathieu}, {Donar}, \&
  {Dullighan}}]{jensen2004}
{Jensen}, E.~L.~N., {Mathieu}, R.~D., {Donar}, A.~X., \& {Dullighan}, A. 2004,
  \apj, 600, 789

\bibitem[{{J{\o}rgensen} {et~al.}(2007){J{\o}rgensen}, {Bourke}, {Myers}, {Di
  Francesco}, {van Dishoeck}, {Lee}, {Ohashi}, {Sch{\"o}ier}, {Takakuwa},
  {Wilner}, \& {Zhang}}]{jorgensen2007}
{J{\o}rgensen}, J.~K., {Bourke}, T.~L., {Myers}, P.~C., {et~al.} 2007, \apj,
  659, 479

\bibitem[{{J{\o}rgensen} {et~al.}(2009){J{\o}rgensen}, {van Dishoeck},
  {Visser}, {Bourke}, {Wilner}, {Lommen}, {Hogerheijde}, \&
  {Myers}}]{jorgensen2009}
{J{\o}rgensen}, J.~K., {van Dishoeck}, E.~F., {Visser}, R., {et~al.} 2009,
  \aap, 507, 861

\bibitem[{{Kenyon} \& {Hartmann}(1995)}]{kenyon1995}
{Kenyon}, S.~J. \& {Hartmann}, L. 1995, \apjs, 101, 117

\bibitem[{{Kratter} {et~al.}(2010){Kratter}, {Matzner}, {Krumholz}, \&
  {Klein}}]{kratter2010}
{Kratter}, K.~M., {Matzner}, C.~D., {Krumholz}, M.~R., \& {Klein}, R.~I. 2010,
  \apj, 708, 1585

\bibitem[{{Kraus} \& {Hillenbrand}(2009)}]{kraus2009}
{Kraus}, A.~L. \& {Hillenbrand}, L.~A. 2009, \apj, 704, 531

\bibitem[{{Launhardt} {et~al.}(2013){Launhardt}, {Stutz}, {Schmiedeke},
  {Henning}, {Krause}, {Balog}, {Beuther}, {Birkmann}, {Hennemann},
  {Kainulainen}, {Khanzadyan}, {Linz}, {Lippok}, {Nielbock}, {Pitann}, {Ragan},
  {Risacher}, {Schmalzl}, {Shirley}, {Stecklum}, {Steinacker}, \&
  {Tackenberg}}]{launhardt2013}
{Launhardt}, R., {Stutz}, A.~M., {Schmiedeke}, A., {et~al.} 2013, \aap, 551,
  A98

\bibitem[{{Lee} {et~al.}(2010){Lee}, {Hasegawa}, {Hirano}, {Palau}, {Shang},
  {Ho}, \& {Zhang}}]{lee2010}
{Lee}, C.-F., {Hasegawa}, T.~I., {Hirano}, N., {et~al.} 2010, \apj, 713, 731

\bibitem[{{Lee} {et~al.}(2009){Lee}, {Hirano}, {Palau}, {Ho}, {Bourke},
  {Zhang}, \& {Shang}}]{lee2009}
{Lee}, C.-F., {Hirano}, N., {Palau}, A., {et~al.} 2009, \apj, 699, 1584

\bibitem[{{Lee} {et~al.}(2016){Lee}, {Dunham}, {Myers}, {Arce}, {Bourke},
  {Goodman}, {J{\o}rgensen}, {Kristensen}, {Offner}, {Pineda}, {Tobin}, \&
  {Vorobyov}}]{lee2016}
{Lee}, K.~I., {Dunham}, M.~M., {Myers}, P.~C., {et~al.} 2016, \apjl, 820, L2

\bibitem[{{Lee} {et~al.}(2015){Lee}, {Dunham}, {Myers}, {Tobin}, {Kristensen},
  {Pineda}, {Vorobyov}, {Offner}, {Arce}, {Li}, {Bourke}, {J{\o}rgensen},
  {Goodman}, {Sadavoy}, {Chandler}, {Harris}, {Kratter}, {Looney}, {Melis},
  {Perez}, \& {Segura-Cox}}]{lee2015}
{Lee}, K.~I., {Dunham}, M.~M., {Myers}, P.~C., {et~al.} 2015, \apj, 814, 114

\bibitem[{{Looney} {et~al.}(2000){Looney}, {Mundy}, \& {Welch}}]{looney2000}
{Looney}, L.~W., {Mundy}, L.~G., \& {Welch}, W.~J. 2000, \apj, 529, 477

\bibitem[{{Maercker} {et~al.}(2012){Maercker}, {Mohamed}, {Vlemmings},
  {Ramstedt}, {Groenewegen}, {Humphreys}, {Kerschbaum}, {Lindqvist},
  {Olofsson}, {Paladini}, {Wittkowski}, {de Gregorio-Monsalvo}, \&
  {Nyman}}]{maercker2012}
{Maercker}, M., {Mohamed}, S., {Vlemmings}, W.~H.~T., {et~al.} 2012, \nat, 490,
  232

\bibitem[{{McCabe} {et~al.}(2006){McCabe}, {Ghez}, {Prato}, {Duch{\^e}ne},
  {Fisher}, \& {Telesco}}]{mccabe2006}
{McCabe}, C., {Ghez}, A.~M., {Prato}, L., {et~al.} 2006, \apj, 636, 932

\bibitem[{{Monin} {et~al.}(2006){Monin}, {M{\'e}nard}, \&
  {Peretto}}]{monin2006}
{Monin}, J.-L., {M{\'e}nard}, F., \& {Peretto}, N. 2006, \aap, 446, 201

\bibitem[{{Murillo} \& {Lai}(2013)}]{murillo2013}
{Murillo}, N.~M. \& {Lai}, S.-P. 2013, \apjl, 764, L15

\bibitem[{{Offner} {et~al.}(2010){Offner}, {Kratter}, {Matzner}, {Krumholz}, \&
  {Klein}}]{offner2010}
{Offner}, S.~S.~R., {Kratter}, K.~M., {Matzner}, C.~D., {Krumholz}, M.~R., \&
  {Klein}, R.~I. 2010, \apj, 725, 1485

\bibitem[{{Offner} {et~al.}(2012){Offner}, {Robitaille}, {Hansen}, {McKee}, \&
  {Klein}}]{offner2012b}
{Offner}, S.~S.~R., {Robitaille}, T.~P., {Hansen}, C.~E., {McKee}, C.~F., \&
  {Klein}, R.~I. 2012, \apj, 753, 98

\bibitem[{{Palau} {et~al.}(2014){Palau}, {Zapata}, {Rodr{\'{\i}}guez}, {Bouy},
  {Barrado}, {Morales-Calder{\'o}n}, {Myers}, {Chapman}, {Ju{\'a}rez}, \&
  {Li}}]{palau2014}
{Palau}, A., {Zapata}, L.~A., {Rodr{\'{\i}}guez}, L.~F., {et~al.} 2014, \mnras,
  444, 833

\bibitem[{{Pezzuto} {et~al.}(2012){Pezzuto}, {Elia}, {Schisano}, {Strafella},
  {Di Francesco}, {Sadavoy}, {Andr{\'e}}, {Benedettini}, {Bernard}, {di
  Giorgio}, {Facchini}, {Hennemann}, {Hill}, {K{\"o}nyves}, {Molinari},
  {Motte}, {Nguyen-Luong}, {Peretto}, {Pestalozzi}, {Polychroni}, {Rygl},
  {Saraceno}, {Schneider}, {Spinoglio}, {Testi}, {Ward-Thompson}, \&
  {White}}]{pezzuto2012}
{Pezzuto}, S., {Elia}, D., {Schisano}, E., {et~al.} 2012, \aap, 547, A54

\bibitem[{{Plunkett} {et~al.}(2013){Plunkett}, {Arce}, {Corder}, {Mardones},
  {Sargent}, \& {Schnee}}]{plunkett2013}
{Plunkett}, A.~L., {Arce}, H.~G., {Corder}, S.~A., {et~al.} 2013, \apj, 774, 22

\bibitem[{{Poglitsch} {et~al.}(2010){Poglitsch}, {Waelkens}, {Geis},
  {Feuchtgruber}, {Vandenbussche}, {Rodriguez}, {Krause}, {Renotte}, {van
  Hoof}, {Saraceno}, {Cepa}, {Kerschbaum}, {Agn{\`e}se}, {Ali}, {Altieri},
  {Andreani}, {Augueres}, {Balog}, {Barl}, {Bauer}, {Belbachir}, {Benedettini},
  {Billot}, {Boulade}, {Bischof}, {Blommaert}, {Callut}, {Cara}, {Cerulli},
  {Cesarsky}, {Contursi}, {Creten}, {De Meester}, {Doublier}, {Doumayrou},
  {Duband}, {Exter}, {Genzel}, {Gillis}, {Gr{\"o}zinger}, {Henning},
  {Herreros}, {Huygen}, {Inguscio}, {Jakob}, {Jamar}, {Jean}, {de Jong},
  {Katterloher}, {Kiss}, {Klaas}, {Lemke}, {Lutz}, {Madden}, {Marquet},
  {Martignac}, {Mazy}, {Merken}, {Montfort}, {Morbidelli}, {M{\"u}ller},
  {Nielbock}, {Okumura}, {Orfei}, {Ottensamer}, {Pezzuto}, {Popesso},
  {Putzeys}, {Regibo}, {Reveret}, {Royer}, {Sauvage}, {Schreiber}, {Stegmaier},
  {Schmitt}, {Schubert}, {Sturm}, {Thiel}, {Tofani}, {Vavrek}, {Wetzstein},
  {Wieprecht}, \& {Wiezorrek}}]{poglitsch2010}
{Poglitsch}, A., {Waelkens}, C., {Geis}, N., {et~al.} 2010, \aap, 518, L2

\bibitem[{{Raghavan} {et~al.}(2010){Raghavan}, {McAlister}, {Henry}, {Latham},
  {Marcy}, {Mason}, {Gies}, {White}, \& {ten Brummelaar}}]{raghavan2010}
{Raghavan}, D., {McAlister}, H.~A., {Henry}, T.~J., {et~al.} 2010, \apjs, 190,
  1

\bibitem[{{Rebull}(2015)}]{rebull2015}
{Rebull}, L.~M. 2015, \aj, 150, 17

\bibitem[{{Reipurth}(2000)}]{reipurth2000}
{Reipurth}, B. 2000, \aj, 120, 3177

\bibitem[{{Reipurth} \& {Mikkola}(2012)}]{reipurth2012}
{Reipurth}, B. \& {Mikkola}, S. 2012, \nat, 492, 221

\bibitem[{{Reipurth} {et~al.}(2010){Reipurth}, {Mikkola}, {Connelley}, \&
  {Valtonen}}]{reipurth2010}
{Reipurth}, B., {Mikkola}, S., {Connelley}, M., \& {Valtonen}, M. 2010, \apjl,
  725, L56

\bibitem[{{Robitaille} {et~al.}(2006){Robitaille}, {Whitney}, {Indebetouw},
  {Wood}, \& {Denzmore}}]{robitaille2006}
{Robitaille}, T.~P., {Whitney}, B.~A., {Indebetouw}, R., {Wood}, K., \&
  {Denzmore}, P. 2006, \apjs, 167, 256

\bibitem[{{Rodr{\'{\i}}guez} {et~al.}(1997){Rodr{\'{\i}}guez}, {Anglada}, \&
  {Curiel}}]{rodriguez1997}
{Rodr{\'{\i}}guez}, L.~F., {Anglada}, G., \& {Curiel}, S. 1997, \apjl, 480,
  L125

\bibitem[{{Rodr{\'{\i}}guez} {et~al.}(1999){Rodr{\'{\i}}guez}, {Anglada}, \&
  {Curiel}}]{rodriguez1999}
{Rodr{\'{\i}}guez}, L.~F., {Anglada}, G., \& {Curiel}, S. 1999, \apjs, 125, 427

\bibitem[{{Sandell} \& {Knee}(2001)}]{sandell2001}
{Sandell}, G. \& {Knee}, L.~B.~G. 2001, \apjl, 546, L49

\bibitem[{{Soderblom} {et~al.}(2014){Soderblom}, {Hillenbrand}, {Jeffries},
  {Mamajek}, \& {Naylor}}]{soderblom2014}
{Soderblom}, D.~R., {Hillenbrand}, L.~A., {Jeffries}, R.~D., {Mamajek}, E.~E.,
  \& {Naylor}, T. 2014, Protostars and Planets VI, 219

\bibitem[{{Stamatellos} \& {Whitworth}(2009{\natexlab{a}})}]{stamatellos2009a}
{Stamatellos}, D. \& {Whitworth}, A.~P. 2009{\natexlab{a}}, \mnras, 392, 413

\bibitem[{{Stamatellos} \& {Whitworth}(2009{\natexlab{b}})}]{stamatellos2009b}
{Stamatellos}, D. \& {Whitworth}, A.~P. 2009{\natexlab{b}}, \mnras, 400, 1563

\bibitem[{{Tafalla} {et~al.}(2006){Tafalla}, {Kumar}, \&
  {Bachiller}}]{tafalla2006}
{Tafalla}, M., {Kumar}, M.~S.~N., \& {Bachiller}, R. 2006, \aap, 456, 179

\bibitem[{{Tobin} {et~al.}(2015){Tobin}, {Dunham}, {Looney}, {Li}, {Chandler},
  {Segura-Cox}, {Sadavoy}, {Melis}, {Harris}, {Perez}, {Kratter},
  {J{\o}rgensen}, {Plunkett}, \& {Hull}}]{tobin2015}
{Tobin}, J.~J., {Dunham}, M.~M., {Looney}, L.~W., {et~al.} 2015, \apj, 798, 61

\bibitem[{{Tobin} {et~al.}(2016){Tobin}, {Looney}, {Li}, {Chandler}, {Dunham},
  {Segura-Cox}, {Sadavoy}, {Melis}, {Harris}, {Kratter}, \&
  {Perez}}]{tobin2016}
{Tobin}, J.~J., {Looney}, L.~W., {Li}, Z.-Y., {et~al.} 2016, \apj, 818, 73

\bibitem[{{van Kempen} {et~al.}(2009){van Kempen}, {van Dishoeck},
  {G{\"u}sten}, {Kristensen}, {Schilke}, {Hogerheijde}, {Boland}, {Menten}, \&
  {Wyrowski}}]{vankempen2009}
{van Kempen}, T.~A., {van Dishoeck}, E.~F., {G{\"u}sten}, R., {et~al.} 2009,
  \aap, 507, 1425

\bibitem[{{Velusamy} \& {Langer}(1998)}]{velusamy1998}
{Velusamy}, T. \& {Langer}, W.~D. 1998, \nat, 392, 685

\bibitem[{{Walawender} {et~al.}(2005){Walawender}, {Bally}, {Kirk}, \&
  {Johnstone}}]{walawender2005}
{Walawender}, J., {Bally}, J., {Kirk}, H., \& {Johnstone}, D. 2005, \aj, 130,
  1795

\bibitem[{{Walawender} {et~al.}(2006){Walawender}, {Bally}, {Kirk},
  {Johnstone}, {Reipurth}, \& {Aspin}}]{walawender2006}
{Walawender}, J., {Bally}, J., {Kirk}, H., {et~al.} 2006, \aj, 132, 467

\bibitem[{{Whitney} {et~al.}(2003){Whitney}, {Wood}, {Bjorkman}, \&
  {Wolff}}]{whitney2003}
{Whitney}, B.~A., {Wood}, K., {Bjorkman}, J.~E., \& {Wolff}, M.~J. 2003, \apj,
  591, 1049

\bibitem[{{Yen} {et~al.}(2015){Yen}, {Koch}, {Takakuwa}, {Ho}, {Ohashi}, \&
  {Tang}}]{yen2015}
{Yen}, H.-W., {Koch}, P.~M., {Takakuwa}, S., {et~al.} 2015, \apj, 799, 193

\bibitem[{{Y{\i}ld{\i}z} {et~al.}(2015){Y{\i}ld{\i}z}, {Kristensen}, {van
  Dishoeck}, {Hogerheijde}, {Karska}, {Belloche}, {Endo}, {Frieswijk},
  {G{\"u}sten}, {van Kempen}, {Leurini}, {Nagy}, {P{\'e}rez-Beaupuits},
  {Risacher}, {van der Marel}, {van Weeren}, \& {Wyrowski}}]{yildiz2015}
{Y{\i}ld{\i}z}, U.~A., {Kristensen}, L.~E., {van Dishoeck}, E.~F., {et~al.}
  2015, \aap, 576, A109

\bibitem[{{Young} {et~al.}(2015){Young}, {Young}, {Lai}, {Dunham}, \&
  {Evans}}]{young2015}
{Young}, K.~E., {Young}, C.~H., {Lai}, S.-P., {Dunham}, M.~M., \& {Evans}, II,
  N.~J. 2015, \aj, 150, 40

\end{thebibliography}

\begin{appendix}
\section{HIPE map makers and photometry}
\label{app:maps}
The three HIPE map-makers (High Pass Filter; Jscan map; MADmap) were tested to determine the best map for performing photometry. The test was made only on the 70 $\mu$m maps for Perseus. The method used involved performing aperture photometry on the source and the surrounding background at four positions. The aperture (12") used was the same for all source and background measurements. Ten sources were selected from different regions of Perseus, ranging from isolated to clustered sources. 

The flux was calculated in the following manner:
\begin{displaymath}
F_{source} = (F'_{source} - \frac{B_1 + ... + B_n}{n}) A_{\rm corr}
\end{displaymath}
where $F'_{source}$ is the raw flux, $B_{i}$ is the background flux, $A_{\rm corr}$ is the aperture correction factor, and $F_{source}$ is the background corrected flux. Aperture correction values were taken from Balog et al. (2014). For an aperture of 12", the correction factor is of 0.802.

Table~\ref{tabres} lists the background and aperture corrected results. This shows that the difference between maps is not significant. We have adopted JScanmap for photometry.

\begin{table*}
\centering
\caption{Aperture photometry results for 70 $\mu$m}
\begin{tabular}{cccccccc}
\hline
Source & RA & Dec & HPF & JSM & MAD & Average & Std. Dev. \\
\hline
IRAS4C & 03:29:13.81 & 31:13:56.11 & 2.72 & 2.74 & 2.75 & 2.74 & 0.014 \\
SK1 & 03:29:00.77 & 31:11:57.59 & 1.42 & 1.43 & 1.37 & 1.41 & 0.026 \\
NGC1333 S1 & 03:28:45.40 & 31:05:40.30 & 1.25 & 1.23 & 1.20 & 1.22 & 0.021 \\
NGC1333 S2 & 03:28:34.49 & 31:00:50.20 & 3.22 & 3.20 & 3.11 & 3.18 & 0.049 \\
IRAS03282 & 03:31:20.99 & 30:45:28.48 & 6.20 & 6.18 & 6.04 & 6.14 & 0.069 \\
IRAS03292 & 03:32:17.95 & 30:49:46.46 & 2.55 & 2.55 & 2.47 & 2.52 & 0.039 \\
L1448IRS1 & 03:25:09.54 & 30:46:20.80 & 2.62 & 2.61 & 2.52 & 2.58 & 0.044 \\
IRAS5 Per63 & 03:28:43.54 & 31:17:31.61 & 2.01 & 2.04 & 1.96 & 2.00 & 0.035 \\
IC348 & 03:33:27.40 & 31:07:10.00 & 4.28 & 4.35 & 4.24 & 4.29 & 0.049 \\
L1455-FIR N & 03:27:38.44 & 30:13:57.95 & 2.23 & 2.25 & 2.33 & 2.27 & 0.043 \\
\hline
\end{tabular}
\label{tabres}
\end{table*}

\section{Evolutionary stage classification}
\label{app:class}
\begin{figure}
\includegraphics[width=\columnwidth]{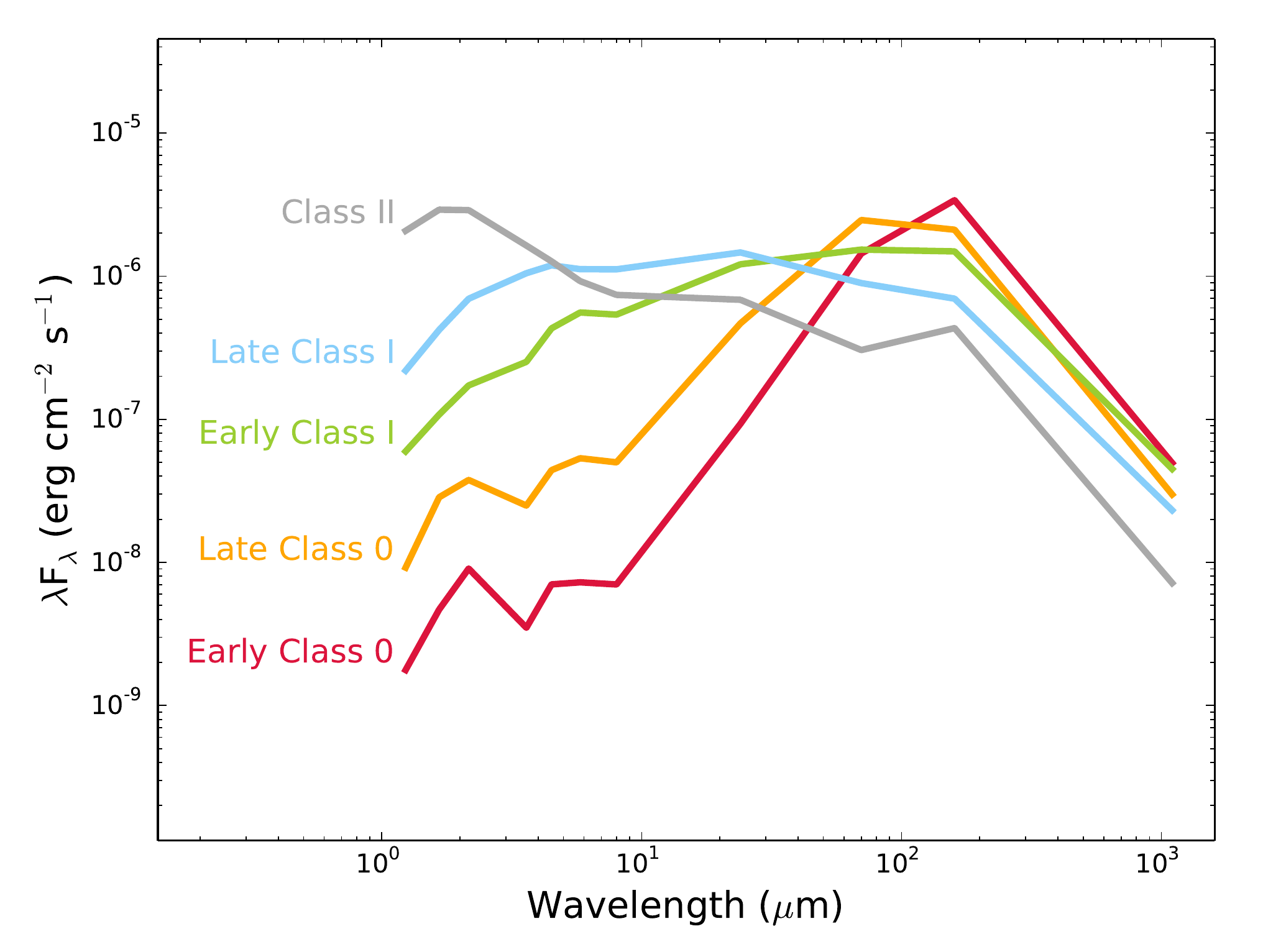}
\caption{Average SEDs derived by \citet{enoch2009}. The classifications are defined based on the bolometric temperature. These average SEDs are used to compare to the constructed SEDs in this work.}
\label{fig:avgsed}
\end{figure}

The SEDs of the resolved systems are compared to average SEDs obtained by (\citealt{enoch2009}; Fig.~\ref{fig:avgsed}) to determine by eye whether the multiple systems are coeval or not.

\begin{figure}
\includegraphics[width=\columnwidth]{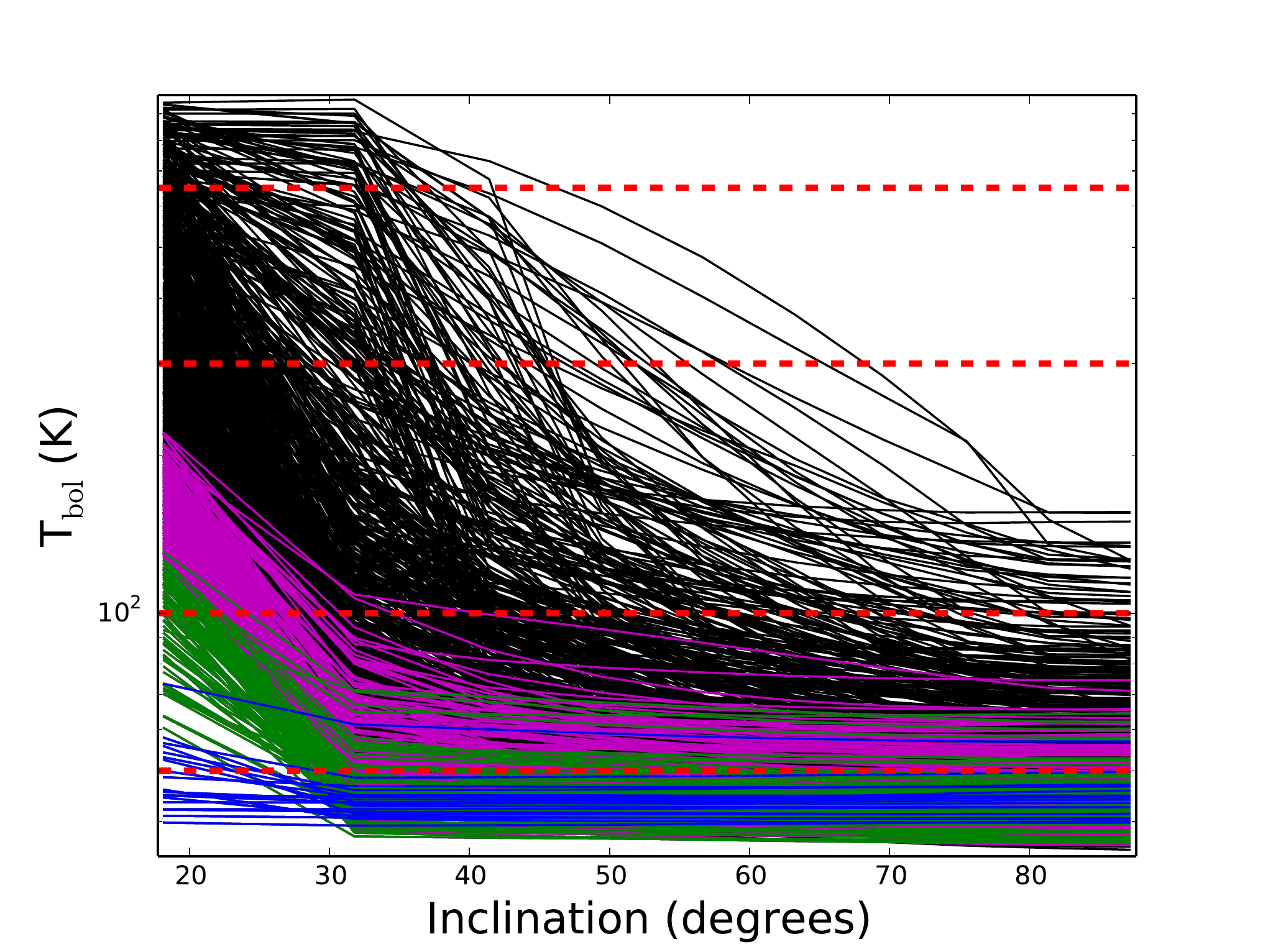}
\includegraphics[width=\columnwidth]{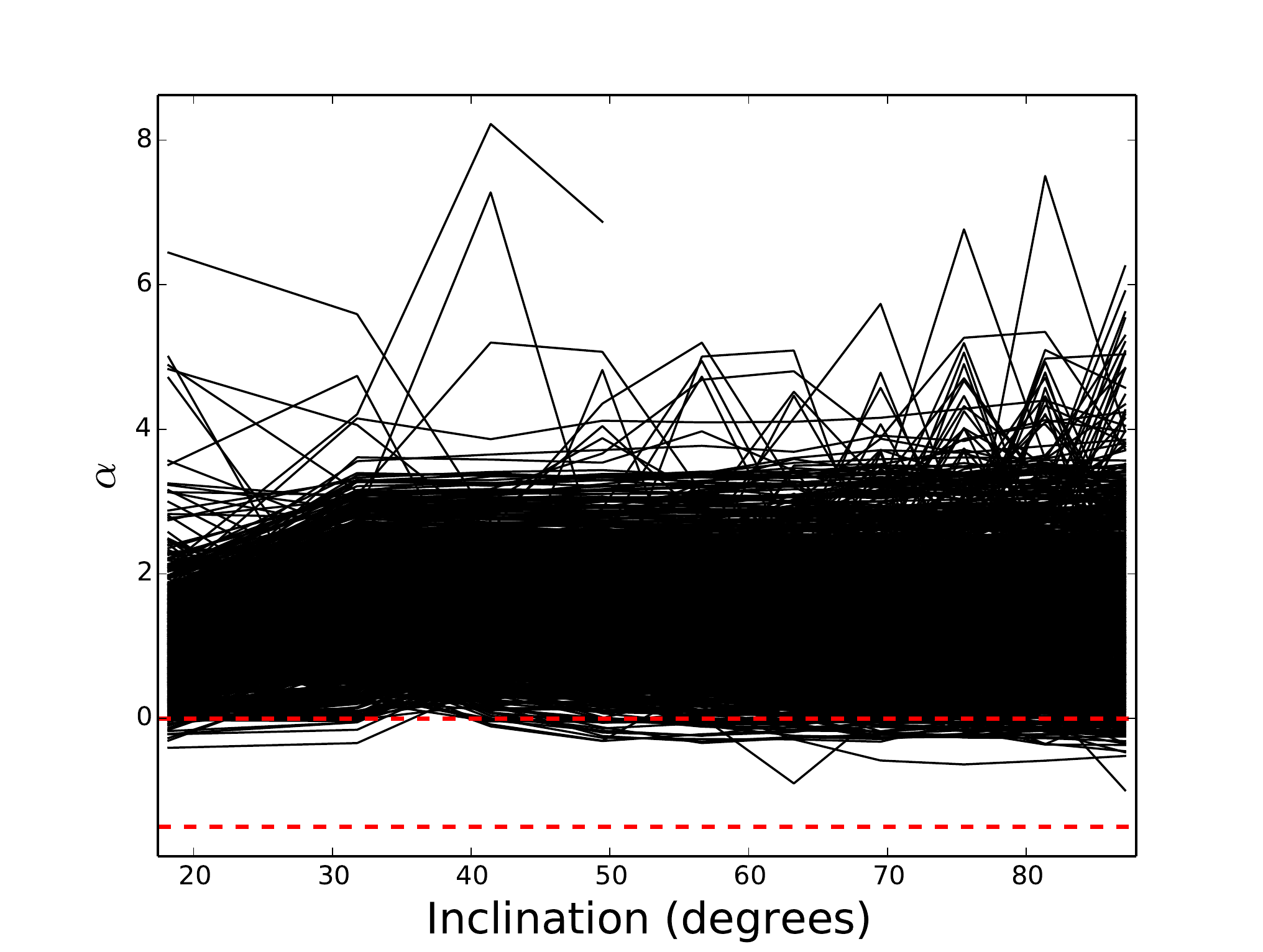}
\includegraphics[width=\columnwidth]{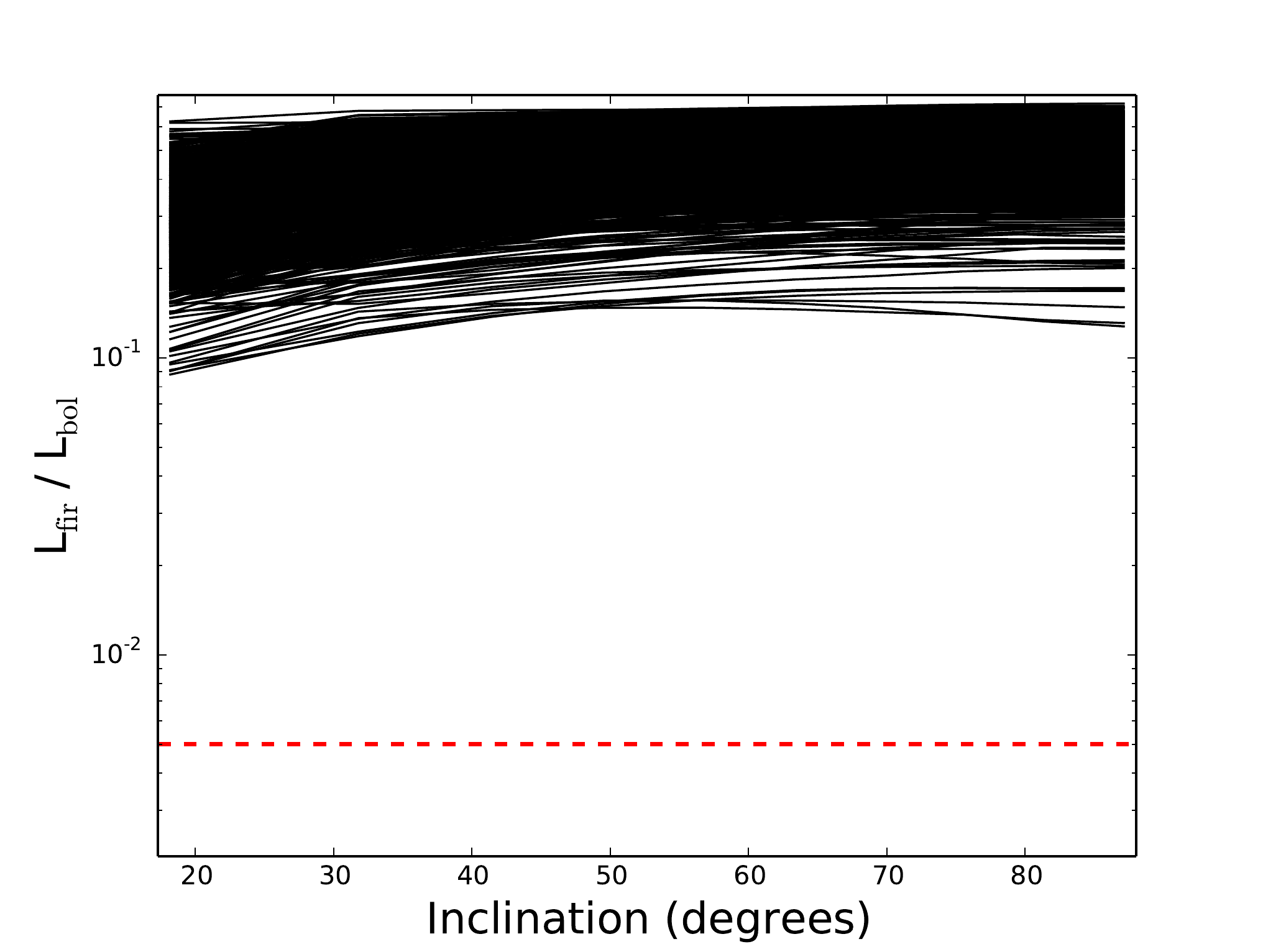}
\caption{Physical parameters derived from the model SEDs from the grid of \citet{robitaille2006} compared to the inclination w.r.t. line of sight. Edge-on is defined as 90$^{\circ}$. The dashed red lines indicate the evolutionary stage boundaries. \textit{Top:} Bolometric temperature $T_{\rm bol}$ varies largely as it inclines more toward face-on (0$^{\circ}$). The blue lines are models with a small difference in $T_{\rm bol}$ with inclination; these have dense envelopes. Green and magenta show the models with an increase in $T_{\rm bol}$ when the inclination is close to face-on. \textit{Middle:} The infrared spectral index from 2 to 24 $\mu$m oscillates without relation to inclination. This is most likely an effect of the ice feature at 8 $\mu$m. \textit{Bottom:} The luminosity ratio $L_{\rm fir} / L_{\rm bol}$ is independent of inclination, but there is no clear boundary for systems at different evolutionary stages.}
\label{fig:modelparams}
\end{figure}

Physical parameters derived from the SED are known to be sensitive to inclination. 
This is confirmed by the properties derived from the model SEDs.
The derived properties were calculated for the model SEDs in the same way as for the observed SEDs.
Comparing the derived parameters, we find that $T_{\rm bol}$ varies widely with inclination, while $L_{\rm fir} / L_{\rm bol}$ are independent of inclination. These results confirm previous work on the subject \citep{jorgensen2009,launhardt2013}.
Figure~\ref{fig:modelparams} shows the parameters for all the Class 0 models versus inclination.

\section{Resolved multiple systems}
\label{app:systems}

\subsection{NGC1333}
\label{appsub:ngc1333}
\noindent\textbf{NGC1333 SVS13:} Located at the heart of NGC1333, SVS13 is a quintuple system with components SVS13A1 and 2, VLA3, SVS13B and C \citep{tobin2016}. 
Based on the velocity field, VLA3 and SVS13B are suggested to be a binary even though VLA3 is closer to SVS13A \citep{chen2009}.
One of the components in SVS13A is expected to be the driving source of HH 7-11 \citep{rodriguez1997,looney2000}. 
SVS13A is observed to have a prominent molecular outflow in the SE-NW direction with a moderate inclination angle and wide opening angles \citep{plunkett2013}. 
These outflow characteristics together with the presence of a centimeter source and being the exciting source for HH 7-11 \citep{rodriguez1997} suggest that SVS13A is a Class 0/I transition object.
SVS13C may be driving a N-S outflow possibly directed along the plane of the sky, but its outflow emission may be confused with other outflows \citep{plunkett2013}.
SVS13B does not show a clear outflow, which could be due to confusion with the outflow of SVS13A \citep{plunkett2013}.
SVS13 has been suggested many times to be non-coeval because components SVS13B and C are more embedded than SVS13A, which has an optical counterpart \citep{looney2000,chen2009}.
SVS13B is not detected in the \textit{Herschel} PACS 70$\mu$m maps, but is detected at $\lambda \geq$100$\mu$m, suggesting it is deeply embedded.

\noindent\textbf{NGC1333 IRAS2:} located in the west of the NGC1333 region, the IRAS2 system is composed of sources A, B and C \citep{sandell2001}, although source C is expected to be a starless core.
IRAS2B is known to be confused with a field star at $\lambda \leq$8$\mu$m \citep{rodriguez1999}.
IRAS2A is typically classified as Class 0, while IRAS2B is considered Class I, but here we find them to be coeval.
IRAS2A is well known because of its spectacular quadrupole outflow. The N-S outflow has a shell-like structure, while the E-W outflow is more collimated \citep{plunkett2013}.
\citep{tobin2015} resolved components with a separation of 0.6$\arcsec$ toward IRAS2A and suggested that the southern component drives the E-W outflow, while the northern component drives the N-S outflow.
The outflow of IRAS2B runs parallel to the N-S outflow of IRAS2A
\citep{plunkett2013}.

\noindent\textbf{NGC1333 IRAS7:} three systems in a common core (CLASSy N$_{2}$H$^{+}$ observations) of about 30$\arcsec$ diameter. Per49 is located in the SE edge of the dust and gas core, while Per18 is located at the peak with Per21 13$\arcsec$ to the SW.
Per18 and Per49 were found to be close binaries with separations $\leq$0.3$\arcsec$ \citep{tobin2016}.
This system of sources is associated with the Haro-Herbig object HH6.
The outflow of this system has been less frequently observed, with candidate outflow lobes proposed to extend to around the NE of the SVS13A outflow \citep{plunkett2013}.
$^{12}$CO observations with CARMA show outflow lobes associated with Per21, but no clear outflow signatures toward the other sources.

\noindent\textbf{NGC1333 IRAS4:} the region contains several systems and IRAS4B' can be resolved at the resolution of Herschel, but IRAS4A cannot.
IRAS4B' (also referred to as IRAS4C, \citealt{looney2000}, and IRAS4B2, \citealt{hull2014}) is not detected in the Herschel observations, which may mean that it is still deeply embedded.
\citet{hull2014} detected outflows toward both sources, with IRAS4B showing an N-S outflow and IRAS4B' driving a weak E-W outflow.

\noindent\textbf{NGC1333 IRAS5:} located at the western edge of NGC1333, this protobinary composed of Per52 and Per63 does not show a prominent molecular outflow \citep{curtis2010}.

\noindent\textbf{NGC1333 Per58+Per65:} located to the north of the NGC1333 core, there are indications of outflow from these sources, although their orientations are unclear because of confusion \citep{curtis2010}.

\noindent\textbf{NGC1333 Per37:} located along a filament in the northernmost region of NGC1333, this triple protostellar system, identified by \citep{tobin2016}, does not appear to have a strong outflow in the observations of \citet{curtis2010}. 

\subsection{L1448 and L1455}
\label{appsub:l1448}

\noindent\textbf{L1448 C:} is a protobinary in the south of L1448, with the southern component having a projected position in the outflow of the northern source.
L1448 C-N shows a more prominent high-velocity collimated outflow traced in $^{12}$CO and SiO, with a low velocity conic cavity observed in $^{12}$CO, while L1448 C-S shows a much weaker low velocity $^{12}$CO collimated outflow \citep{hirano2010}.
Both outflows are not aligned, but do not show signs of significant misalignment along the line of sight either.

\noindent\textbf{L1448 N:} also commonly known as L1448 IRS3, this system is a sextuple, with the B and C sources containing three and two components, respectively \citep{lee2015}. 
All three sources drive molecular outflows, with components B and C almost parallel to each other and source A perpendicular to B and C \citep{lee2015}. The observed outflows suggest no significant misalignment along the line of sight.
The outflow from L1448 C was suggested to induce fragmentation in this core \citep{barsony1998}.

\noindent\textbf{L1448 IRS2:} \citet{tobin2016} found L1448 IRS2 to be a close binary (separation $\sim$0.7$\arcsec$). SCUBA 850$\mu$m observations showed a continuum peak to the east of L1448 IRS2, which \citet{chen2010} referred to as IRS2E, and together with SMA and \textit{Spitzer} upper limits proposed to be a first core candidate. Although SCUBA 850$\mu$m observations suggest a shared envelope, the separation of these two sources is 46$\arcsec$ and therefore cannot compose a multiple system.
The outflow of IRS2 is observed to be conical in the SE-NW direction, while the suggested outflow of IRS2E is composed of only the red-shifted lobe directed toward the SW \citep{hull2014,chen2010}.

\subsection{IC348}
\label{appsub:ic348}

\noindent\textbf{IC348 Per8+Per55:} this triple protostellar system shows a large-scale jet directed in the north-south direction and is associated with two Haro-Herbig objects, HH841 and HH842 \citep{walawender2006}. Based on the large-scale jet, there is no indication of multiple components.

\noindent\textbf{IC348 MMS:} a protobinary located in the southwest of IC348 and associated with a strong north-south outflow where the overlapping of redshifted emission at the tip of the blueshifted lobe is suggested to be due to a change in environment and not a product of inclination; see \citet{tafalla2006}.
The outflow is driven by the Class 0 western source, MMS2, and is also associated with HH797. 
The eastern source, MMS2E, was suggested to be a Class 0 proto-brown dwarf driving a weak outflow in the NE-SW direction by \citet{palau2014}.
Both sources are found to be coeval, as suggested by \citet{palau2014}.

\noindent\textbf{IC348 SMM2:} also referred to as Per16+Per28, this protobinary shows jet emission in the east-west direction for both components and a large-scale S-shape bend in the flow \citep{walawender2006}. The outflows appear to be short and clumpy \citep{eisloffel2003,walawender2006}.

\noindent\textbf{IC348 Per32+EDJ2009-366:} the H$_{2}$ emission around these sources appears to be clumpy, with bow shocks pointing east and possibly not all clumps belonging to the respective sources \citep{eisloffel2003}. Per32 is a close protobinary ($\sim$6$\arcsec$) \citep{tobin2016} and was also found to be a low luminosity object \citep{dunham2008,hsieh2015}.

\subsection{B-1}
\label{appsub:b1}

\noindent\textbf{B1 Per6+Per10:} the outflows of this protobinary are not well determined owing to their location with several neighbouring systems whose outflows become entangled \citep{walawender2005,hiramatsu2010}. Per10 might drive a red-shifted outflow lobe directed NW, but it is uncertain whether the lobe belongs to Per10 or to a source about $\sim$70$\arcsec$ to the north \citep{hiramatsu2010}. Per6 is also known as SMM3 and Per10 as SSTc2d J033314.4+310711.

\noindent\textbf{B1-b:} a triple system composed of Per41 to the west and B1-bN and B1-bS to the north and south, respectively. 
B1-bN and S are found to be very young based on 7 to 1.1 mm continuum data and lack of \textit{Spitzer} detections at $\lambda \leq$24$\mu$m \citep{hirano2014}.
The outflows for both sources are directed E-W, but the location of the blueshifted lobes suggests that the two sources are misaligned, since the north source has the outflow directed to the west, while the south source has it directed to the east \citep{gerin2015}.
Per41 appears to be older, with no detection in the millimeter regime. It is suggested that there is no emission in the millimeter  because the envelope of this system is being striped off by neighboring outflows, which makes it appear to be more evolved \citep{hirano2014}.

\section{Single protostellar systems}
\label{app:singles}
\begin{figure*}
\includegraphics[width=\textwidth]{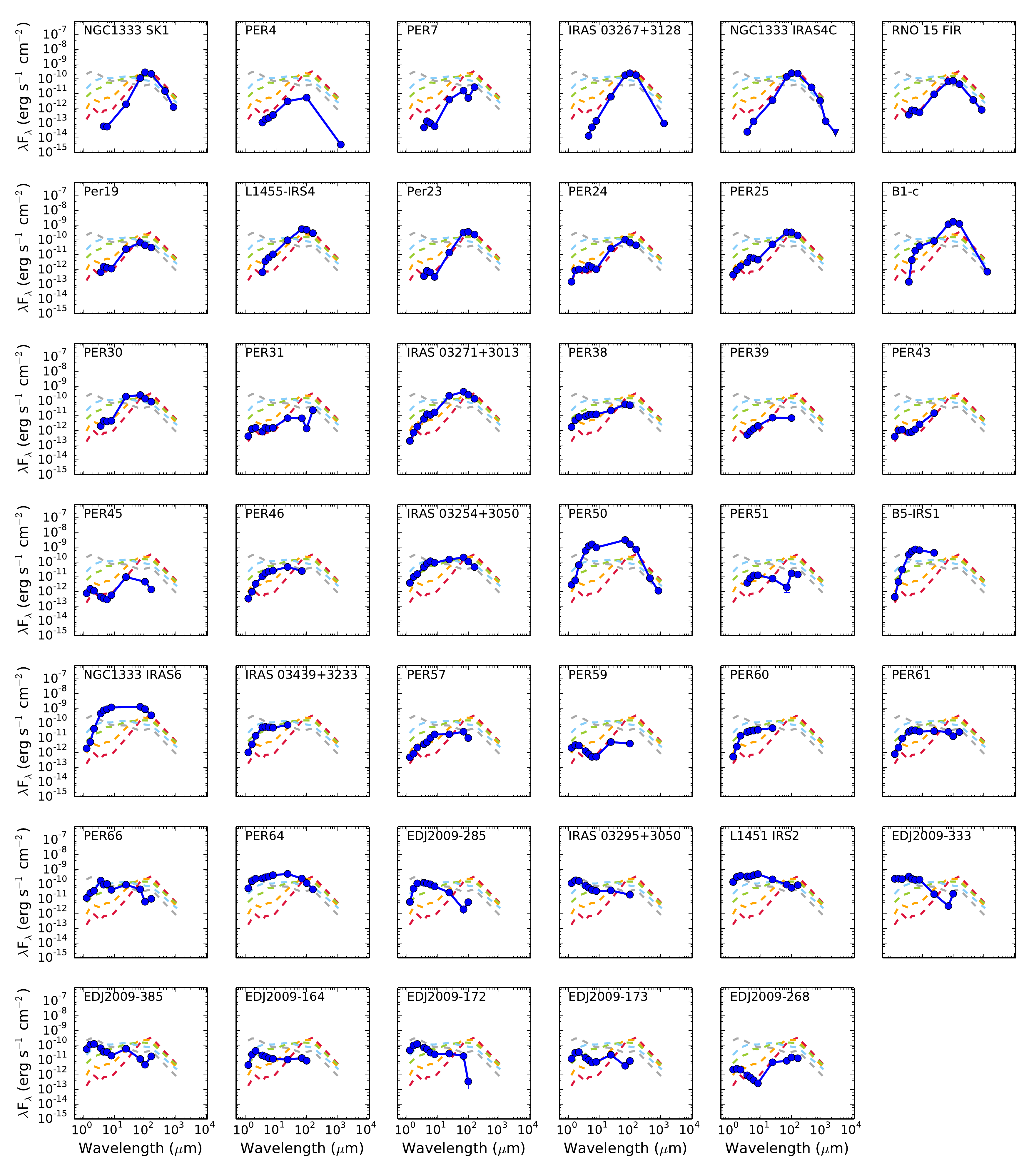}
\caption{Constructed SEDs for single protostellar systems. Other details are the same as in Figure~\ref{fig:SEDresolved}.}
\label{fig:SEDsingle}
\end{figure*}

Single protostellar sources indentified in Perseus \citep{tobin2016} are listed in Table~\ref{tab:single} and the constructed SEDs are shown in Fig.~\ref{fig:SEDsingle}.
Not all sources have an SED, either due to lack of fluxes in the literature, non-detection in the \textit{Herschel} PACS maps, or both. 
This is denoted in the last column of Table~\ref{tab:single}.

\begin{table*}
\caption{Sample of single protostellar systems}
\label{tab:single}
\centering
\begin{tabular}{c c c c}
\hline \hline
Source & RA & Dec. & Constructed SED? \\
\hline
NGC1333 SK1 & 03:29:00.52 & +31:12:00.68 & Y \\
PER4 & 03:28:39.10 & +31:06:01.80 & Y \\
PER7 & 03:30:32.68 & +30:26:26.48 & Y \\
IRAS 03267+3128 & 03:29:51.82 & +31:39:06.08 & Y \\
NGC1333 IRAS4C & 03:29:13.52 & +31:13:58.01 & Y \\
RNO 15 FIR & 03:29:04.05 & +31:14:46.61 & Y \\
Per19 & 03:29:23.49 & +31:33:29.48 & Y \\
L1455-IRS4 & 03:27:43.23 & +30:12:28.80 & Y \\
Per23 & 03:29:17.16 & +31:27:46.41 & Y \\
PER24 & 03:28:45.30 & +31:05:41.99 & Y \\
PER25 & 03:26:37.46 & +30:15:28.01 & Y \\
B1-c & 03:33:17.85 & +31:09:32.00 & Y \\
PER30 & 03:33:27.28 & +31:07:10.20 & Y \\
PER31 & 03:28:32.55 & +31:11:05.21 & Y \\
IRAS 03271+3013 & 03:30:15.12 & +30:23:49.20 & Y \\
PER38 & 03:32:29.18 & +31:02:40.88 & Y \\
PER39 & 03:33:13.78 & +31:20:05.21 & Y \\
PER43 & 03:42:02.16 & +31:48:02.09 & Y \\
PER45 & 03:33:09.57 & +31:05:31.20 & Y \\
PER46 & 03:28:00.40 & +30:08:01.28 & Y \\
IRAS 03254+3050 & 03:28:34.50 & +31:00:51.09 & Y \\
PER50 & 03:29:07.76 & +31:21:57.21 & Y \\
PER51 & 03:28:34.53 & +31:07:05.49 & Y \\
B5-IRS1 & 03:47:41.56 & +32:51:43.89 & Y \\
NGC1333 IRAS6 & 03:29:01.57 & +31:20:20.69 & Y \\
IRAS 03439+3233 & 03:47:05.42 & +32:43:08.41 & Y \\
PER57 & 03:29:03.33 & +31:23:14.60 & Y \\
PER59 & 03:28:35.04 & +30:20:09.89 & Y \\
PER60 & 03:29:20.07 & +31:24:07.49 & Y \\
PER61 & 03:44:21.33 & +31:59:32.60 & Y \\
PER66 & 03:43:45.15 & +32:03:58.61 & Y \\
PER64 & 03:33:12.85 & +31:21:24.08 & Y \\
PER-BOLO-58 & 03:29:25.46 & +31:28:14.99 & N \\ 
PER-BOLO-45 & 03:29:07.70 & +31:17:16.80 & N \\ 
L1451-MMS & 03:25:10.21 & +30:23:55.20 & N \\ 
IRAS 03363+3207 & 03:39:25.20 & +32:17:03.29 & N \\ 
EDJ2009-161 & 03:28:51.48 & +30:45:00.48 & N \\ 
EDJ2009-263 & 03:30:27.45 & +30:28:27.43 & N \\ 
EDJ2009-285 & 03:32:46.94 & +30:59:17.80 & Y \\
IRAS 03295+3050 & 03:32:34.15 & +31:00:56.22 & Y \\
L1451 IRS2 & 03:27:47.49 & +30:12:05.32 & Y \\
EDJ2009-333 & 03:42:55.77 & +31:58:44.39 & Y \\
EDJ2009-385 & 03:44:17.91 & +32:04:57.08 & Y \\
EDJ2009-164 & 03:28:53.96 & +31:18:09.35 & Y \\
EDJ2009-172 & 03:28:56.65 & +31:18:35.44 & Y \\
EDJ2009-173 & 03:28:56.96 & +31:16:22.20 & Y \\
SVS3 & 03:29:10.42 & +31:21:59.07 & N \\
EDJ2009-268 & 03:30:38.23 & +30:32:11.67 & Y \\
\hline
\end{tabular}
\end{table*}

\section{\textit{Herschel} Catalog}
\label{app:fluxes}
This appendix contains the \textit{Herschel} PACS flux catalog for the Perseus star forming region obtained in this work through PSF photometry with \textit{StarFinder} \citep{diolaiti2000}.
The fluxes have been background and aperture corrected, with the aperture correction values tabulated in \citet{balog2014}.

\onecolumn
\begin{longtable}{ccccccccc}
\caption{\textit{Herschel} PACS protostellar fluxes for Perseus}\\
\hline\hline
Source & RA & Dec. & $F_{\rm 70, int}$ & $F_{\rm 70, err}$ & $F_{\rm 100, int}$ & $F_{\rm 100, err}$ & $F_{\rm 160, int}$ & $F_{\rm 160, err}$ \\
 & & & mJy & mJy & mJy & mJy & mJy & mJy \\
\hline
\endfirsthead
\caption{continued.}\\
\hline\hline
Source & RA & Dec. & $F_{\rm 70, int}$ & $F_{\rm 70, err}$ & $F_{\rm 100, int}$ & $F_{\rm 100, err}$ & $F_{\rm 160, int}$ & $F_{\rm 160, err}$ \\
 & & & mJy & mJy & mJy & mJy & mJy & mJy \\
\hline
\endhead
\hline
\endfoot
L1448 IRS2 & 03:25:22.40 & +30:45:12.00 & 29600.27 & 24.45 & 50315.79 & 9.35 & ... & ... \\
L1448 IRS2E & 03:25:25.66 & +30:44:56.70 & 120.00 & 0.00 & ... & ... & 2700.00 & 0.00 \\
L1448N C & 03:25:35.53 & +30:45:34.20 & 5975.70 & 25.02 & 15971.03 & 8.10 & 34348.59 & 12.01 \\
L1448N B & 03:25:36.34 & +30:45:14.94 & 10245.52 & 45.09 & 35063.42 & 10.66 & ... & ... \\
L1448N A & 03:25:36.53 & +30:45:21.35 & 33206.75 & 31.12 & 56855.07 & 10.66 & 133706.90 & 12.01 \\
L1448C N & 03:25:38.87 & +30:44:05.40 & 68675.92 & 33.84 & 86927.75 & 9.23 & 105056.08 & 11.74 \\
L1448C S & 03:25:39.14 & +30:43:58.30 & 10598.25 & 32.11 & 20902.32 & 8.74 & ... & ... \\
Per25 & 03:26:37.46 & +30:15:28.01 & 7640.36 & 24.54 & 10878.83 & 9.43 & 10587.80 & 11.71 \\
L1455 FIR2 & 03:27:38.23 & +30:13:58.80 & 3040.50 & 24.57 & 3573.99 & 9.88 & 6011.33 & 17.46 \\
Per17 & 03:27:39.09 & +30:13:03.00 & 70668.57 & 24.45 & 80705.92 & 9.29 & 58554.50 & 11.72 \\
L1455-IRS4 & 03:27:43.23 & +30:12:28.80 & 12583.17 & 31.19 & 15966.15 & 9.12 & 15317.03 & 11.69 \\
L1451 IRS2 & 03:27:47.49 & +30:12:05.32 & 2186.14 & 24.36 & 1897.01 & 14.88 & 4605.56 & 16.99 \\
Per46 & 03:28:00.40 & +30:08:01.28 & 575.09 & 24.51 & ... & ... & ... & ... \\
Per31 & 03:28:32.55 & +31:11:05.21 & 155.36 & 24.70 & 45.52 & 7.90 & 1279.57 & 11.78 \\
IRAS 03254+3050 & 03:28:34.50 & +31:00:51.09 & 4738.27 & 24.83 & 3613.08 & 7.93 & 2439.00 & 11.82 \\
Per51 & 03:28:34.53 & +31:07:05.49 & 44.76 & 24.71 & 567.30 & 7.90 & 763.32 & 11.89 \\
Per59 & 03:28:35.04 & +30:20:09.89 & ... & ... & 135.28 & 7.99 & ... & ... \\
Per4 & 03:28:39.10 & +31:06:01.80 & ... & ... & 177.23 & 7.91 & ... & ... \\
NGC1333 IRAS5 Per52 & 03:28:39.72 & +31:17:31.89 & 208.77 & 24.64 & 427.39 & 7.90 & 928.07 & 11.90 \\
NGC1333 IRAS5 Per63 & 03:28:43.28 & +31:17:32.90 & 2631.24 & 24.53 & 1733.94 & 9.54 & 2905.58 & 11.90 \\
Per24 & 03:28:45.30 & +31:05:41.99 & 2440.21 & 24.69 & 2218.64 & 7.92 & 2304.05 & 14.02 \\
EDJ2009-156 & 03:28:51.11 & +31:18:15.41 & 272.70 & 24.57 & 474.27 & 7.90 & ... & ... \\
EDJ2009-164 & 03:28:53.96 & +31:18:09.35 & 316.65 & 24.55 & 297.22 & 7.89 & ... & ... \\
NGC1333 IRAS2A & 03:28:55.57 & +31:14:37.22 & 353085.72 & 24.59 & 457917.62 & 7.91 & 355028.75 & 11.90 \\
NGC1333 Per65 & 03:28:56.31 & +31:22:27.80 & 190.82 & 24.82 & 707.66 & 7.92 & 295.13 & 12.25 \\
EDJ2009-172 & 03:28:56.65 & +31:18:35.44 & 434.22 & 24.67 & 11.57 & 7.91 & ... & ... \\
EDJ2009-173 & 03:28:56.96 & +31:16:22.20 & 95.59 & 24.61 & 287.08 & 7.92 & ... & ... \\
NGC1333 IRAS2B & 03:28:57.35 & +31:14:15.93 & 42904.02 & 24.61 & 42112.11 & 7.90 & 40297.80 & 11.90 \\
NGC1333 Per58 & 03:28:58.44 & +31:22:17.40 & 2023.94 & 24.51 & 1361.60 & 7.92 & 2993.06 & 13.89 \\
EDJ2009-183 & 03:28:59.32 & +31:15:48.14 & 1373.58 & 24.52 & 568.81 & 7.90 & ... & ... \\
NGC1333 SK1 & 03:29:00.52 & +31:12:00.68 & 2639.06 & 24.65 & 9202.65 & 7.93 & 11595.86 & 11.84 \\
NGC1333 IRAS6 & 03:29:01.57 & +31:20:20.69 & 29984.85 & 24.55 & 29552.15 & 7.91 & 18105.12 & 12.75 \\
NGC1333 SVS13C & 03:29:01.96 & +31:15:38.26 & 10414.79 & 28.64 & 25758.22 & 9.50 & 39433.51 & 14.03 \\
NGC1333 SVS13B & 03:29:03.07 & +31:15:52.02 & ... & ... & 29753.50 & 0.00 & 49162.50 & 0.00 \\
Per57 & 03:29:03.33 & +31:23:14.60 & 610.89 & 24.66 & 325.77 & 7.93 & ... & ... \\
NGC1333 SVS13A & 03:29:03.75 & +31:16:03.76 & 346127.50 & 24.55 & 362422.36 & 9.35 & 262763.62 & 13.88 \\
RNO 15 FIR & 03:29:04.05 & +31:14:46.61 & 1557.72 & 24.60 & 2464.90 & 7.91 & 2338.33 & 11.83 \\
Per50 & 03:29:07.76 & +31:21:57.21 & 73021.60 & 24.75 & 53860.12 & 7.90 & 38022.12 & 14.21 \\
NGC1333 IRAS4A & 03:29:10.51 & +31:13:31.01 & 34177.79 & 24.54 & 145231.28 & 7.91 & 198935.09 & 11.79 \\
NGC1333 IRAS7 PER21 & 03:29:10.67 & +31:18:20.18 & 22177.79 & 25.30 & 33424.83 & 8.11 & 32656.71 & 13.73 \\
NGC133 IRAS7 Per18 & 03:29:11.26 & +31:18:31.08 & 29785.00 & 30.59 & 51718.19 & 10.33 & 49503.68 & 14.11 \\
NGC1333 IRAS4B & 03:29:12.01 & +31:13:08.10 & 13967.95 & 24.47 & 57324.07 & 7.92 & 95091.14 & 11.87 \\
NGC1333 IRAS4B' & 03:29:12.83 & +31:13:06.90 & ... & ... & ... & ... & ... & ... \\
NGC1333 IRAS7 Per49 & 03:29:12.96 & +31:18:14.31 & 2287.81 & 24.31 & 1234.96 & 7.93 & ... & ... \\
NGC1333 IRAS4C & 03:29:13.52 & +31:13:58.01 & 3215.11 & 24.49 & 8028.17 & 7.93 & 12204.80 & 11.80 \\
Per23 & 03:29:17.16 & +31:27:46.41 & 7282.82 & 24.53 & 11712.11 & 7.92 & 12160.26 & 11.79 \\
NGC1333 EDJ2009-233 & 03:29:17.675 & +31:22:44.922 & 500.51 & 24.27 & 693.54 & 7.90 & ... & ... \\
NGC1333 EDJ2009-235 & 03:29:18.259 & +31:23:19.758 & ... & ... & ... & ... & ... & ... \\
NGC1333 Per37 & 03:29:18.89 & +31:23:12.89 & 2083.38 & 30.77 & 4311.49 & 7.91 & 8826.94 & 11.79 \\
Per60 & 03:29:20.07 & +31:24:07.49 & ... & ... & ... & ... & ... & ... \\
Per19 & 03:29:23.49 & +31:33:29.48 & 1539.06 & 24.38 & 1437.67 & 7.91 & 1615.74 & 13.53 \\
IRAS 03267+3128 & 03:29:51.82 & +31:39:06.08 & 4111.28 & 24.66 & 7984.09 & 7.92 & 9383.54 & 11.89 \\
IRAS 03271+3013 & 03:30:15.12 & +30:23:49.20 & 9835.80 & 24.40 & 8981.08 & 8.96 & 7599.56 & 13.28 \\
Per7 & 03:30:32.68 & +30:26:26.48 & 367.70 & 24.45 & 166.76 & 7.98 & 1453.21 & 14.87 \\
EDJ2009-268 & 03:30:38.23 & +30:32:11.67 & 206.44 & 24.38 & 501.70 & 7.96 & 704.23 & 11.75 \\
EDJ2009-269 & 03:30:43.91 & +30:32:46.28 & 1281.74 & 24.16 & 588.90 & 7.98 & 969.73 & 14.26 \\
IRAS 03282+3035 & 03:31:21.00 & +30:45:30.00 & 8447.13 & 24.59 & 22504.65 & 9.29 & ... & ... \\
IRAS 03292+3039 & 03:32:17.95 & +30:49:47.60 & 1700.00 & 180.00 & 9100.00 & 1300.00 & 24000.00 & 8400.00 \\
Per38 & 03:32:29.18 & +31:02:40.88 & 1391.34 & 24.58 & 1721.54 & 7.86 & ... & ... \\
IRAS 03295+3050 & 03:32:34.15 & +31:00:56.22 & ... & ... & 645.16 & 7.88 & ... & ... \\
EDJ2009-285 & 03:32:46.94 & +30:59:17.80 & 45.89 & 24.30 & 201.98 & 7.84 & ... & ... \\
Per45 & 03:33:09.57 & +31:05:31.20 & ... & ... & 153.45 & 7.85 & 73.99 & 11.99 \\
Per64 & 03:33:12.85 & +31:21:24.08 & 5666.66 & 24.37 & 3961.29 & 7.83 & 2369.53 & 11.77 \\
Per39 & 03:33:13.78 & +31:20:05.21 & ... & ... & 231.78 & 7.85 & ... & ... \\
NGC1333 PER6 & 03:33:14.40 & +31:07:10.88 & 1259.95 & 24.15 & 1076.28 & 7.87 & 1873.57 & 14.78 \\
NGC1333 PER10 & 03:33:16.45 & +31:06:52.49 & 2514.94 & 24.44 & 6431.20 & 7.86 & 9511.64 & 11.75 \\
B1-a & 03:33:16.66 & +31:07:55.20 & 8844.80 & 24.24 & 7961.24 & 7.86 & ... & ... \\
B1-c & 03:33:17.85 & +31:09:32.00 & 26917.64 & 24.01 & 56567.06 & 7.85 & 65357.92 & 11.83 \\
B1-bW & 03:33:20.30 & +31:07:21.29 & 385.40 & 29.27 & 72.83 & 7.84 & ... & ... \\
B1-bN & 03:33:21.20 & +31:07:44.20 & ... & ... & 356.37 & 7.87 & 2458.19 & 11.87 \\
B1-bS & 03:33:21.30 & +31:07:27.40 & 32.30 & 24.68 & 2169.70 & 7.83 & 9152.34 & 13.65 \\
Per30 & 03:33:27.28 & +31:07:10.20 & 5885.32 & 24.09 & 4870.87 & 7.85 & 4753.57 & 11.81 \\
Per43 & 03:42:02.16 & +31:48:02.09 & ... & ... & ... & ... & ... & ... \\
EDJ2009-333 & 03:42:55.77 & +31:58:44.39 & 76.98 & 28.07 & 751.94 & 8.82 & ... & ... \\
Per66 & 03:43:45.15 & +32:03:58.61 & 1044.54 & 23.67 & 216.00 & 7.51 & 545.83 & 10.67 \\
IC348 SMM2N & 03:43:51.00 & +32:03:23.76 & 1786.13 & 29.32 & 3298.55 & 8.93 & 4712.59 & 10.22 \\
IC348 SMM2S & 03:43:51.08 & +32:03:08.32 & ... & ... & 9069.77 & 7.52 & 8418.18 & 11.32 \\
HH211 & 03:43:56.80 & +32:00:50.21 & 5008.66 & 23.38 & 20474.13 & 7.54 & 29631.02 & 16.35 \\
IC348 MMS Per11 & 03:43:57.06 & +32:03:04.60 & ... & ... & 18485.14 & 8.53 & 22651.04 & 13.02 \\
IC348 MMS2E & 03:43:57.73 & +32:03:10.10 & 360.00 & 50.00 & 1200.00 & 100.00 & 1300.00 & 700.00 \\
EDJ2009-366 & 03:43:59.44 & +32:01:53.99 & 183.88 & 26.69 & 1053.86 & 8.26 & ... & ... \\
Per32 & 03:44:02.40 & +32:02:04.89 & 99.35 & 23.59 & 350.11 & 7.52 & 1506.45 & 10.04 \\
Per62 & 03:44:12.98 & +32:01:35.40 & 7627.83 & 23.53 & 8954.61 & 8.44 & 5590.77 & 10.06 \\
EDJ2009-385 & 03:44:17.91 & +32:04:57.08 & 274.20 & 23.32 & 160.37 & 8.48 & 942.41 & 11.49 \\
PER61 & 03:44:21.33 & +31:59:32.60 & 601.91 & 23.69 & 432.97 & 7.52 & 1326.28 & 10.08 \\
IC348 Per55 & 03:44:43.33 & +32:01:31.41 & 4451.78 & 29.90 & 4821.42 & 7.97 & ... & ... \\
IC348 Per8 & 03:44:43.94 & +32:01:36.09 & 12057.88 & 31.07 & 26087.98 & 8.49 & 28035.05 & 10.09 \\
IRAS 03439+3233 & 03:47:05.42 & +32:43:08.41 & ... & ... & ... & ... & ... & ... \\
B5-IRS1 & 03:47:41.56 & +32:51:43.89 & ... & ... & ... & ... & ... & ... \\
\end{longtable}

\end{appendix}

\end{document}